\documentclass[%
 amsmath,amssymb,
 aps,
pra,
]{revtex4-2}

\usepackage{graphicx}
\usepackage{dcolumn}
\usepackage{bm}

\begin{document}

\preprint{APS/123-QED}

\title{Dead-time optimization to increase secure distance range \\ in prepare and measure quantum key distribution protocols}%

\author{Carlos Wiechers}
\email{e-mail: carherwm@fisica.ugto.mx}
\affiliation{Departamento de F\'isica-DCI, Universidad de Guanajuato, P.O. Box E-143,Le\'on, Gto., 37150, M\'exico.}

\author{X\'ochitl S\'anchez-Lozano}
\affiliation{Universidad de Guanajuato, Lascurain de Retana 5, Zona Centro, Guanajuato, Gto., 36000, M\'exico.}

\author{Rafael G\'omez-Medina}
\affiliation{Divisi\'on de Ciencias e Ingenier\'ias, Campus Le\'on, Universidad de Guanajuato, Loma del Bosque 103, Lomas del Campestre, Le\'on, Gto., 37150, M\'exico.}

\author{Mariana Salado-Mej\'ia}
\affiliation{Instituto Nacional de Astrof\'isica, \'Optica y Electr\'onica (INAOE), Calle Luis Enrique Erro 1, Santa María Tonantzintla, Puebla, 72840, M\'exico}%

\author{J.L. Lucio}
\affiliation{Departamento de F\'isica-DCI, Universidad de Guanajuato, P.O. Box E-143,Le\'on, Gto., 37150, M\'exico.}

\date{\today}

\begin{abstract}
Afterpulsing is a factor limiting the distance over which discrete-variable quantum key distribution systems are secure, and a common feature in single-photon detectors. The relevance of this phenomenon stems from its stochastic, self-interacting nature and the fact that its rate rises with the number of avalanche events, which increases the quantum bit error rate. Here we introduce an effective analytic model, including dead-time and afterpulsing corrections, where afterpulsing correction depends on dead-time value. This model is useful to evaluate the performance of prepare and measure quantum key distribution protocols (standard and decoy versions) that use gated single photon detectors. The model provides an expression to numerically optimize the secret key rate over the full distance range for secure communication, enabling in this way the calculation of quantum bit error rate and secure key rate. In the conventional procedure, the dead-time value is fixed regardless of distance, limiting the distance range of the channel due to remaining afterpulsing effects, which are more relevant at higher operating frequencies. Here we demonstrate that optimizing the dead-time values increases the distance range of the channel to share secret keys. 
\end{abstract}


\maketitle


\section{Introduction}\label{Sec:Int}
We have witnessed a vertiginous technological development thanks to the incorporation of quantum treatment in topics such as information~\cite{Flamini_2018}, metrology~\cite{Giov2011,Xiang2013}, imaging~\cite{Moreau2019}, and sensing~\cite{Degen2017}, making quantum technologies a reality.  Quantum information is one of the most developed of these fields, in particular quantum key distribution (QKD)~\cite{Lut:1,Sca:1,Lo:14,Pira2019}, which allows sharing secure private keys between two authorized parties, taking advantage of physical laws of Quantum Mechanics to secure the transmission of information. Due to quantum nature of the information carriers (signal), if an eavesdropper attempts to obtain information of the key, alters the signal’s state, and leaves a footprint that reveals its activity~\cite{BB84}.  Developments in this field include among others, the well-known Prepare and Measure quantum key distribution (P\&M-QKD) protocols: 1) the standard BB84 protocol proposed by Bennett and Brassard~\cite{BB84}; 2) the standard SARG04 protocol~\cite{Sca:04} proposed as an alternative to overcome the photon-number-splitting (PNS) attack~\cite{Bra00}; 3) and their respective decoy versions~\cite{Hwa03,Lo05,Wan05}, which are even more robust against the PNS attack. These protocols have been extensively studied in the literature, performed in the lab, and are commercially available~\cite{Lo:14, Pira2019}. Although security is guaranteed by physical laws, the performance of actual implementations is limited by equipment and the optimization of its operation parameters.  

Those discrete variable protocols are based on single-photon detectors, such as avalanche photodiodes (APD), superconducting nanowires (SCNW) or photomultiplier tubes (PMT), which have in common, as noise sources, the dark counts and afterpulsing~\cite{Wie:2}. Afterpulsing affects the performance of QKD systems, notably at high operation frequencies, by modifying the raw key generation rate, the quantum bit error rate (QBER), and the amount of quantum error correction in secret key distillation~\cite{Yos:1}. So far afterpulsing effects have been incorporated in the analysis of QKD performance~\cite{Lim2014,Li2015,Fan2018}, and have been the subject of continuous improvements in the modeling~\cite{FanY2020,Wang2022,Huang2022}. In this work we provide also an analytical expression for the afterpulsing and its explicit dependence on dead-time and other timing parameters. 

 In this paper we address the problem of incorporating afterpulsing effects in discrete variable P\&M-QKD systems that rely on the use of single-photon detectors (SPD).  As proof of concept our analysis is performed using APD based SPD, for which the usual dead-time values are in the range of 10 -- 20~$\mu$s. However, in spite of these values, there are still a non-negligible remaining afterpulsing contribution. We introduce a phenomenological recursive model, allowing an analytical description of the afterpulsing contribution. The model incorporates the PNS attack constraints through the mean photon-number value, it includes also afterpulsing and dead-time corrections allowing optimization of the secure key rate and distance range, where the operation of the P\&M-QKD protocols is secure. Details of the model including the standard and decoy versions of the BB84 and SARG04 protocols are discussed in Section~\ref{Sec:2}; while Section~\ref{Sec:4} is devoted to the results of the analyses performed, including comparison with Monte Carlo simulations. Our results are summarized in Section~\ref{Sec:5}. Technical details of the afterpulsing corrections factor is presented in appendix~\ref{Ap:A}. In appendix~\ref{Ap:B}, we report the detector’s parameters fitting using our model.

\begin{figure*}[h!]
\centering
\begin{tabular}{c}
\includegraphics[scale=0.5,trim= 0cm 6cm 3cm 0cm,clip]{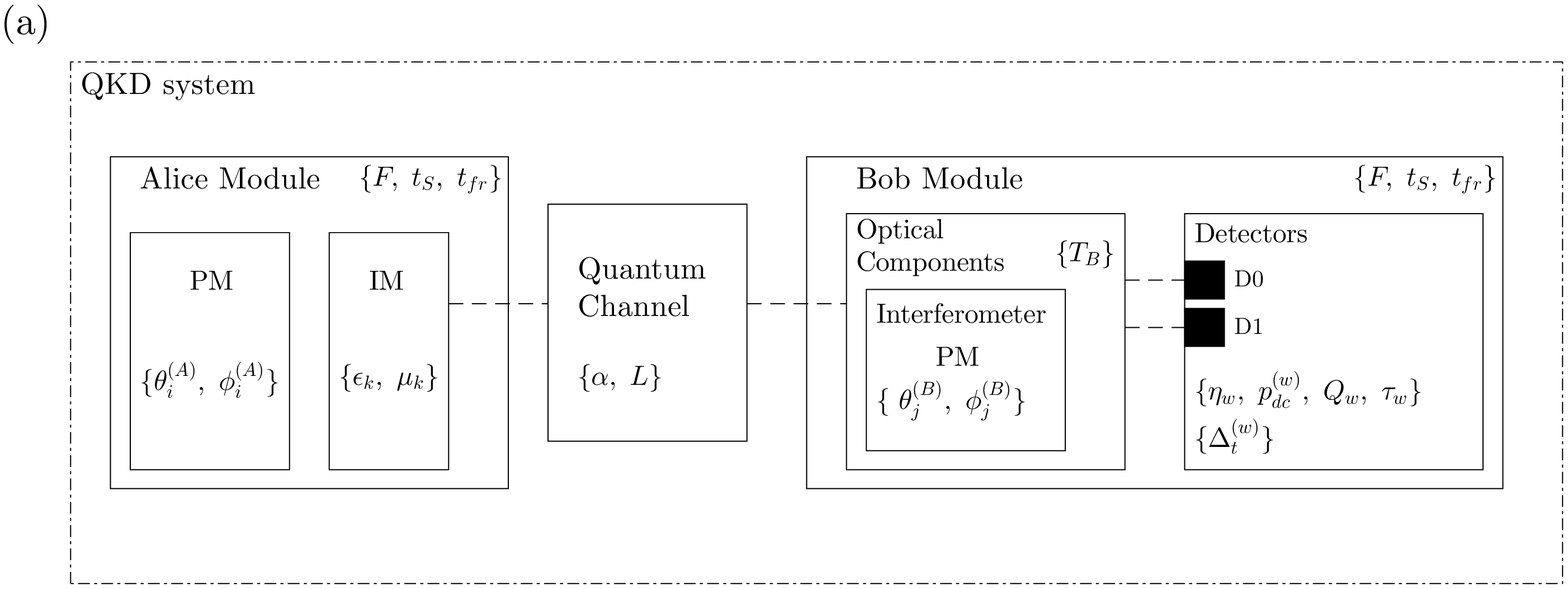}
\\
\includegraphics[scale=0.5,trim= 0cm 2cm 3cm 0cm, clip]{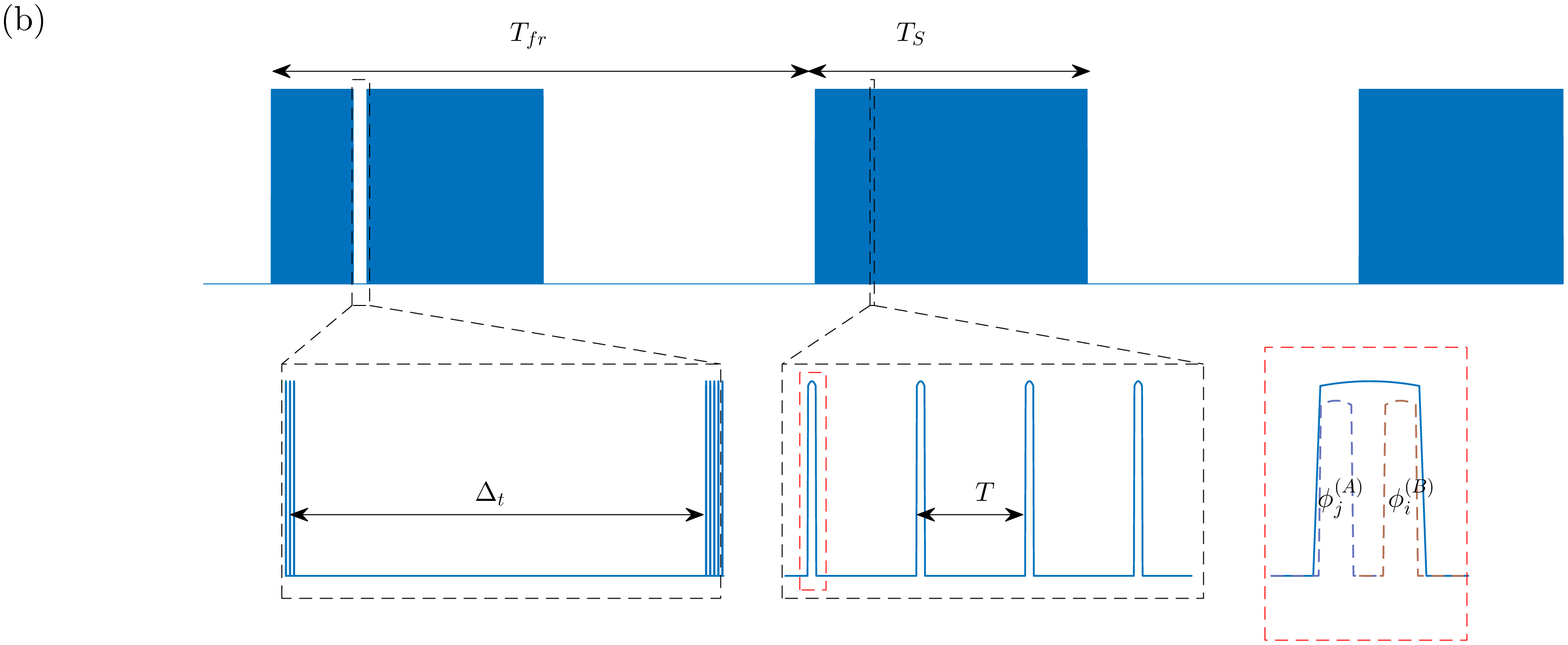}
\end{tabular}
\caption{\small (a) Schematic view of a P\&M-QKD system, showing the set of dynamic parameters, IM stands for intensity modulator and PM$_{A/B}$ for phase modulators. (b) Timing operation parameters: $t_{fr}$ - frame period, $t_S$ - frame duration, $\Delta_t$ - dead-time,  and $t_F=1/F$ - gate period, where $F$ is the operation frequency. The red square we depict a time-bin which is conformed by two pulses signal and reference, where the phases $\phi_i^{(A)}$ and $\phi_i^{(B)}$ are applied.} \label{QKDs}
\end{figure*}

\section{Model} \label{Sec:2}
A schematic, simplified description of a P\&M-QKD system is shown in Fig.~\ref{QKDs}(a). The diagram shows only the dynamic parameters for system operation that are relevant to our discussion. Alice’s module includes the source, a set of weak laser pulses each of those described by a coherent state, which follow a poissonian probability distributions ($P_{n,k}=\mathrm{e}^{-\mu_k}\mu_k^n/n!$), with a mean photon-number $\mu_k$, where $n$ is the number of photons in the pulses. The output-states are prepared using one of $K$ different mean photon-number values, and one of four different phases ($\phi_i^{(A)}$) in the signal pulse and no phase on the reference pulse. The amplitude and phase values are randomly chosen and applied by the intensity modulator (IM$_A$) and phase modulator (PM$_A$), respectively.

Alice sends output-states --signal and reference-- to Bob's Module through the quantum channel (optical fiber), whose effect on the process is described by its attenuation. The transmittance of the channel is given by $T_c=10^{-\alpha L/10}$, where $L$ is the channel length, and $\alpha$ the attenuation coefficient, which for commercial optical fiber has a typical value $\alpha = 0.20$~dB/km.  

In Bob’s module the pulses enter an interferometer. In one of the interferometer paths the reference pulses travel and by means of the phase modulator $PM_B$, one of two phases $\phi_j^{(B)}$ is randomly impressed on them. Meanwhile the signal pulses travel through the other path. Both signal and reference pulses interfere in a final beam splitter and each output ends in a SPD. In this module the pulses undergo an additional attenuation denoted $T_B$, due to the optical components. We use the label $D_w$ with $w\in \{0, 1\}$ to denote SPDs which have the following internal parameters: photodetection efficiency $\eta_w$, dark count probability $p_{dc}^{(w)}$,  and afterpulsing, which is described in terms of the amplitude $Q_w$ and decay time $\tau_w$. The afterpulsing parameters depend on the voltage and temperature settings of the SPDs, and we must keep in mind that $Q_w$ and $\tau_w$ are effective parameters that describe SPDs carrier-trap assembles~\cite{Wie:1}. Additionally, the dead-time $\Delta_t^{(w)}$ is an internal SPDs timing parameter employed to mitigate the afterpulsing contribution to the total detection events. 

In Fig.~\ref{QKDs}(b), we depict $\Delta_t^{(w)}$ and the global external timing parameters: frame of duration $t_{S}$, frame period $t_{fr}$, and gate period $t_F=1/F$, where $F$ is the operating frequency. Also, we include a time-bin (in the red dashed square), which is conformed by two pulses signal and reference. 

\subsection{Total detection probability and correction factors}
In  P\&M QKD protocol SPDs are operated in gated-mode and their detection events are triggered by three processes: photodetection, dark counts, and afterpulsing. Using IM$_A$, Alice prepares one state per laser pulse, choosing randomly one of $K$ different weak coherent states per time beam, each one with a mean photon number $\mu_k$ and a occurrence probability $\epsilon_k$. For the standard BB84 and SARG04 protocols, signal and its reference states have a single coherent state with a mean photon number ($\mu_1$). Considering a common notation in the decoy version of the aforementioned protocols, the coherent states using $\mu_1$ are also for the signal and its reference states, and other coherent states per time beam with  $\mu_{k'}\neq \mu_1$ are used for the decoy-states and its reference state, and satisfy: a) $k' > 1$, b) $\mu_{1} > \sum_{k'=2}^K\mu_{k'}$, c) $\mu_{k'}>\mu_{k'+1}$, and d) $\epsilon_{1} > \sum_{k'=2}^K\epsilon_{k'}$. In our analyses we use $K=3$, for a three-state decoy version. The key is encoded by means of PM$_A$ by printing to each signal pulse one of four different phases $\phi_i^{A}\in\{0,\ \pi/2,\ \pi,\ 3\pi/2 \}$ with occurrence probability $\theta_i^{(A)}$. 

Using PM$_B$ in the interferometer, Bob applies to each signals one of two random phases $\phi_j^{(B)}\in\{0,\ \pi/2 \}$ with occurrence probability $\theta_j^{(B)}$. Then, the probability at each output of the interferometer is given by $\cos^2(\Delta \phi_{ij}-w\pi/2))$, which depends on the phase difference $\Delta \phi_{ij}=(\phi_i^{(A)}-\phi_j^{(B)})/2$, and the value of the label $w$, regarding to each detector. We use the following formula to calculate the average non-detection probability of photodetection events $P_{ph0}^{w}$, where we have considered all possible photodetection outcomes at each detector as follows,
\begin{equation}
P_{ph0}^{(w)}=\sum_{k=1}^K \sum_{i=1}^4 \sum_{j=1}^2 \epsilon_k \theta_i^{(A)}\theta_j^{(B)}\exp \left(-\gamma_{ijw}\mu_k\right), \label{EQ:PH}
\end{equation}
where, $\gamma_{ijw}=\eta_w T_c T_B\cos^2(\Delta \phi_{ij}-w\pi/2))$ is the total detection efficiency considering the internal detector efficiency $\eta_w$, channel transmittance $T_c$, and Bob Module transmittance $T_B$, and interferometric outcomes $\cos^2(\Delta \phi_{ij}-w\pi/2))$. Deviations due to phase fluctuations from the quantum channel or misalignment in Bob Module can be included, using $\Delta \phi_{ij}\rightarrow \Delta \phi_{ij}+\delta$, here we use $\delta=0$.  

We are interested in the detection probability $P_{TC}^{(w)}$ including afterpulsing and dead-time correction factors, which is calculated with,  
\begin{equation}
P_{TC}^{(w)}=P_{T}^{(w)}C^{(w)}.\label{Eq:PTC}
\end{equation}
where the intermediate total detection probability, $P_{T}^{(w)}$, which considers only the afterpulsing correction factor,
\begin{equation}
P_{T}^{(w)}=1-(1-P_{N}^{(w)})P_{ph0}^{(w)}, \label{Eq:PT}
\end{equation}
this probability is required to calculate the dead-time correction factor $(C^{(w)})$, which is described in~\citep{AHun:1,Wie:2}, and its effect is given by,
\begin{equation}
C^{(w)}=\{ P_T^{(w)}(F\Delta_t^{(w)}-1)+1 \} ^{-1}. \label{EQ:C}
\end{equation}

The noise probability, $P_N^{(w)}$, takes into account the afterpulsing correction factor $P_{APC}^{(w)}$ and the dark counts probability $(p_{dc}^{(w)})$, since both of them are the two main sources of noise. It is obtained with, 
\begin{equation}
P_N^{(w)}=\left[1-(1-p_{dc}^{(w)})P_{APC}^{(w)}\right].\label{Eq:PN}
\end{equation}

In general, dead-time does not completely vanish the afterpulsing effects. Therefore, we introduce the afterpulsing correction factor, we regard as the afterpulsing core probability ($P_{APC}^{(w)}$), representing the average non-detection probability of the entire set of afterpulsing events and is expressed as,
\begin{equation}
P_{APC}^{(w)}=(1-P_{TC}^{(w)}P_{af}^{(w)})^{N_a^{(w)}},  \label{EQ:APC}
\end{equation}
where, $P_{af}^{(w)}$ is the average afterpulsing probability per gate; the exponent $N_a^{(w)}=(N^{(w)}+1)/2$ is the average number of gates per frame;  $N^{(w)}=(t_S-\Delta_t^{(w)})F$ is the maximum number of gates affected by afterpulsing contribution, $t_S$ is the frame duration, $\Delta_t^{(w)}$ is the detector dead-time, and $F$ the operation frequency. $P_{APC}^{(w)}$ considers the cumulative effects due to afterpulsing from previous detection events. The afterpulsing probability $p_{AP}$ is given by,
\begin{equation}
    p_{AP}^{(w)}=1-P_{APC}^{(w)}. \label{Eq:PAP}
\end{equation}
Let us now return to the determination of the average afterpulsing probability per gate $P_{af}^{(w)}$. To start with notice that although the physics behind  afterpulsing is different for each type of detector, the phenomenological description using exponential decay function is universal. Thus, we assume the afterpulsing probability is described by an  effective decay model with an unique decay time. Afterpulsing probability increases with the number of previous events, leading its cumulative nature, we use the formulation derived in Appendix~\ref{Ap:A} to obtain the average afterpulsing probability contribution per gate, 
\begin{equation} 
P_{af}^{(w)}=\beta_w k_wQ_w\exp\left( -k_w\Delta_t^{(w)} \right),\label{Eq:DA}
\end{equation}
where, $k_w=\tau_w^{-1}$, is the decay rate, $k_wQ_w\exp\left( -k_w\Delta_t^{(w)} \right)$ is the afterpulsing probability induced by a detection event mitigated after a single dead-time. $\beta_w$ is a factor regarding the average of the afterpulsing  contribution, and its analytic expression is   
\begin{eqnarray}
\beta_w = \frac{1}{N_a^{(w)}\left[1-\rho_w\right]} \left\lbrace 1-\frac{\rho_w\left[1-\rho_w^{N^{(w)}}\right]}{N^{(w)}\left[1-\rho_w\right]}\right\rbrace, \label{Eq:be}
\end{eqnarray}
where, $\rho_w = \mathrm{e}^{-k_w t_F}$. 

Since of $P_T^{(w)}$ is necessary to calculate both correction factors ( $C^{(w)}$ and $P_{APC}^{(w)}$) and is difficult to invert. We use 
\begin{equation}
    P_{T,(0)}^{(w)}=1-(1-p_{dc}^{(w)})P_{ph0}^{(w)}, 
\end{equation}
as the initial value in the recursive calculation until the value numerically converges to a small value $\varepsilon$ as,
\begin{equation}
    \varepsilon>\left|\frac{P_{T,(l+1)}^{(w)}-P_{T,(l)}^{(w)}}{P_{T,(l)}^{(w)}}\right|.
\end{equation}

\subsection{Sifted key rate, QBER and secret key}
The (P\&M-QKD) protocols rely on the use of four different phases $\phi_i^{(A)}$ to prepare the signal and two possible phases $\phi_j^{(B)}$ at Bob's interferometer,  leading to a total of eighth phase combinations in $\Delta\phi_{ij}$ prior to detection. The probability of detection of an actual signal $p_{i,\mu_k}^{(w)}$, using the correction factors, is calculated using the signal state which correspond to those labeled  with $\mu_1$. Of the eight possible phase combinations, we show four of them using a single basis in Bob, the other four results using the second basis are equivalent, below we only report the four results for a single basis, 
\begin{eqnarray}
p_{1,\mu_1}^{(w)}&=&C^{(w)}\left[1-\left(1-p_{N}^{(w)}\right)\mathrm{e}^{-\Gamma_w \mu_1} \right],\label{Eq:p1}\\
p_{2,\mu_1}^{(w)}&=&C^{(w)}p_{N}^{(w)},\label{Eq:p2}\\
p_{3,\mu_1}^{(w)}&=&C^{(w)}\left[1-\left(1-p_{N}^{(w)}\right)\mathrm{e}^{-\Gamma_w \mu_1 /2} \right],\label{Eq:p3}\\
p_{4,\mu_1}^{(w)}&=&p_{3,\mu_1}^{(w)},\label{Eq:p4} 
\end{eqnarray}
these probabilities are obtained from Eq.~\ref{EQ:PH}, using a $k=1$, and different values of $i$ and a single value of $j$. $\Gamma_w=\eta_wT_cT_B$ is the reduced total efficiency. We introduce $h_{ijw}=\Delta \phi_{ij}-w\pi/2$, as the phase argument in the square cosine function of $\gamma_{ijw}$ in Eq.~\ref{EQ:PH}. The values $h_{ijw}=0$, and $h_{ijw}=\pi/2$, lead to deterministic outcomes given in Eqs.~\ref{Eq:p1}-\ref{Eq:p2}, respectively. Meanwhile, Eqs.~\ref{Eq:p3}-\ref{Eq:p4} are the non-deterministic outcomes with $h_{ijw}=\pm\pi/4\ \mathrm{or}\ 3\pi/4$. These equations use $w\in \{0,\ 1\}$ as SPDs identifier. Additionally, the dead-time correction factor, Eq.~\ref{EQ:C}, and noise probability, E.~\ref{Eq:PN}, which includes the afterpulsing correction factor.  

At this point, we use equations~\ref{Eq:p1}-\ref{Eq:p4} to remove the possible double counts among both detectors, so that we end up with the probability of single events of signals (without double counts),
\begin{eqnarray}
P_{1,\mu_1}^{(w)}&=& p_{1,\mu_1}^{(w)}\left[1-p_{2,\mu_1}^{(1-w)}\right],\label{Eq:P1}\\
P_{2,\mu_1}^{(w)}&=& p_{2,\mu_1}^{(w)}\left[1-p_{1,\mu_1}^{(1-w)}\right],\label{Eq:P2}\\
P_{3,\mu_1}^{(w)}&=& p_{3,\mu_1}^{(w)}\left[1-p_{4,\mu_1}^{(1-w)}\right].\label{Eq:P3}
\end{eqnarray}
For P\&M protocols, the sifted key rate ($R_{\mu_1}$) and total QBER ($E_{\mu_1}$) are given in terms of the probability of single detection events of signals,
\begin{eqnarray}
R_{\mu_1}&=& \sum_{w=0}^1 \left( P_{q,\mu_1}^{(w)}+P_{2,\mu_1}^{(w)}\right),    \label{EQ:RT}\\
E_{\mu_1}&=& \frac{1}{R_{\mu_1}}\sum_{w=0}^1 P_{2,\mu_1}^{(w)}    , \label{EQ:ET}
\end{eqnarray}
where here and thereafter, $q=1$ for BB84 and BB84 decoy; and $q=3$ for SARG04 and SARG04 decoy. We present the results of the decoy method, where  afterpulsing and dead-time correction factors are considered. We transform the probabilities from Eqs.~\ref{Eq:p1}-\ref{Eq:p4} into following the proto-yield equations, 
\begin{eqnarray}
z_{1,n}^{(w)}&=&C^{(w)}\left[1-\left(1-p_{N}^{(w)}\right)\left(1-\Gamma_w \right)^n \right],\label{Eq:z1}\\
z_{2,n}^{(w)}&=&C^{(w)}p_{N}^{(w)},\label{Eq:z2}\\
z_{3,n}^{(w)}&=&C^{(w)}\left[1-\left(1-p_{N}^{(w)}\right)\left(1-\Gamma_w/2\right)^n \right],\label{Eq:z3}\\
z_{4,n}^{(w)}&=&z_{3,n}^{(w)},\label{Eq:z4}
\end{eqnarray}
where $n$ is number of photon in the field. A direct connection among the total probability and proto-yield equations is given by $p_{m,\mu_1}^{(w)}=\sum_{n=0}^\infty z_{m,n}^{(w)}\mathrm{e}^{-\mu_1}\mu_1^n/n!$. If we use the following values $C^{(w)}=1$ and $P_{APC}^{(w)}=1$, we recover the original results of the decoy method~\cite{Hwa03,Lo05,Wan05}.\\
To remove the double counts among the detectors in the proto-yield Eqs~\ref{Eq:z1}-\ref{Eq:z4}, we introduce the following proto-yield probabilities of single detection events,
\begin{eqnarray}
Z_{1,n}^{(w)}&=& z_{1,n}^{(w)}\left[1-p_{2,\mu_1}^{(1-w)}\right],\label{Eq:Z1}\\
Z_{2,n}^{(w)}&=& z_{2,n}^{(w)}\left[1-p_{1,\mu_1}^{(1-w)}\right],\label{Eq:Z2}\\
Z_{3,n}^{(w)}&=& z_{3,n}^{(w)}\left[1-p_{4,\mu_1}^{(1-w)}\right].\label{Eq:Z3}
\end{eqnarray}
The n-photon yields ($Y_n$), their corresponding bit error ($e_n$) and rates ($r_n$) are calculated with Eqs.~\ref{Eq:Z1}-\ref{Eq:Z3},
\begin{eqnarray}
Y_{n}&=&\sum_{w=0}^1 \left( Z_{q,n}^{(w)}+Z_{2,n}^{(w)}\right),     \label{EQ:yn}\\
r_n&=& Y_n\mu_1^n\mathrm{e}^{-\mu_1}/n!,      \label{EQ:rn}\\
e_{n}&=&\frac{1}{Y_{n}}\sum_{w=0}^1 Z_{2,n}^{(w)}.     \label{EQ:en}
\end{eqnarray}
In this work, we consider random phase assignment, \textit{i.e.} no bias is included, for which we use $\theta_i^{(A)}=1/4$ and $\theta_j^{(B)}=1/2$ for all values of $i$ and $j$, respectively. We use the results of Eq.~\ref{EQ:RT} and Eq.~\ref{EQ:ET}, together with  Eq.~\ref{EQ:rn} and Eq.~\ref{EQ:en} with $n=1$, to obtain the secret key rate, 
\begin{eqnarray}
S&=&\frac{\epsilon_1N^{(w)}}{2t_{fr}} [ r_1\left(1-\mathrm{H}_2\left( e_1\right)\right) \nonumber\\
 & &-f\left(E_{\mu_1}\right)R_{\mu_1}\mathrm{H}_2\left(E_{\mu_1}\right) ], \label{EQ:S}
\end{eqnarray}
with $\mathrm{H}_2(u)=-u\log_2(u)-(1-u)\log_2(1-u)$, the Shannon's entropy, and $t_{fr}$ is the repetition frame period. Only the single photon states are considered secure. Error correction leads to data loss, $f$ stands for the error correction factor,  and the term $f\left(E_{\mu_1}\right)R_{\mu_1}\mathrm{H}_2\left(E_{\mu_1}\right)$ quantifies the lost data.

\section{Results} \label{Sec:4}
In our approach, the problem amounts to determine values of the internal detector parameters entering Eq.~\ref{EQ:S}. The procedure to  achieve this is the following: 
We assume that  the voltage and temperature settings on the SPDs are fixed, therefore we first characterize the SPDs. In Table~\ref{T:1}, we report the values of the internal detector parameters that  we use for our analyses. Meanwhile, the optimal value of $\mu_1$ is obtained by limiting the PNS attack~\cite{Bra00,Lut02}; for the standard BB84 and SARG04 protocols, $\mu_1$ depends on the channel transmittance $T_c$, $\mu_1=T_c$ and $\mu_1=\sqrt{2}T_c^{1/2}$, respectively. In the case of their decoy state versions, the criteria is to maximize the secret key rate and minimize the PNS attack, considering the single-photon component.\\
In the following sub-section we present the optimization process and then perform a Monte Carlo simulation to compare with the model results.

\begin{table}[ht!]
\centering
\caption{\small Internal detector parameters. The following values are obtained from a fit analysis of Fig~\ref{F:9}. Fitting error: $\sigma_e=0.0214$}\label{T:1}
\begin{tabular}{ccccc}
\hline
\hline
 $\eta/10^{-2}$ &   $p_{dc}/10^{-5}$  &   $Q$ (ns)&   $\tau$ ($\mu$s)  \\
\hline
 $9.32\pm 0.20$ & $2.028\pm0.009$ & $15.35\pm 0.13$& $71.5\pm 4.2$  \\
\hline
\hline
\end{tabular}
\end{table}

\subsection{Optimization of P\&M-QKD protocols}
\noindent Our model involves a large number of parameters that can be used to optimize the performance of QKD protocols. However, it is not necessary to consider all of them in the optimization. In fact we know that the secret key rate $S$ increases when $t_S$, or $\epsilon_1$ increase; and $S$ decreases if $t_{fr}-t_S$, or $\alpha$, increase. We also know that $S$ decreases if $\epsilon_2$, $\epsilon_3$, $\mu_2$ or $\mu_3$ increase. In the optimization process we keep fixed the parameters related to the decoy states, since they do not contribute to the secret key rate, but generate afterpulsing. We also keep fixed the internal parameters of the detector, since we choose a configuration of the SPDs and keep it.

\begin{figure}[h!]
\centering
\begin{tabular}{cc}
\includegraphics[scale=0.35]{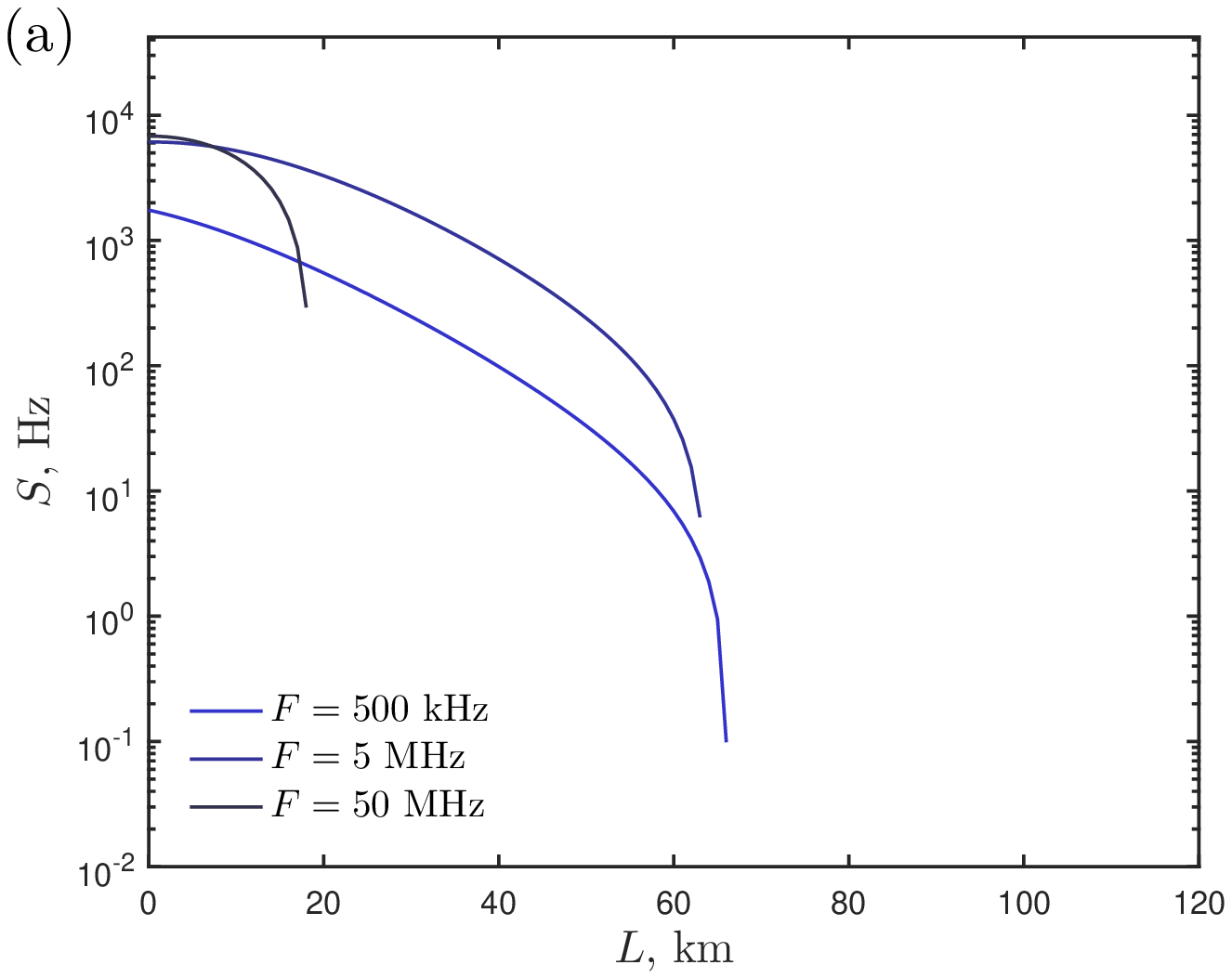}&\includegraphics[scale=0.35]{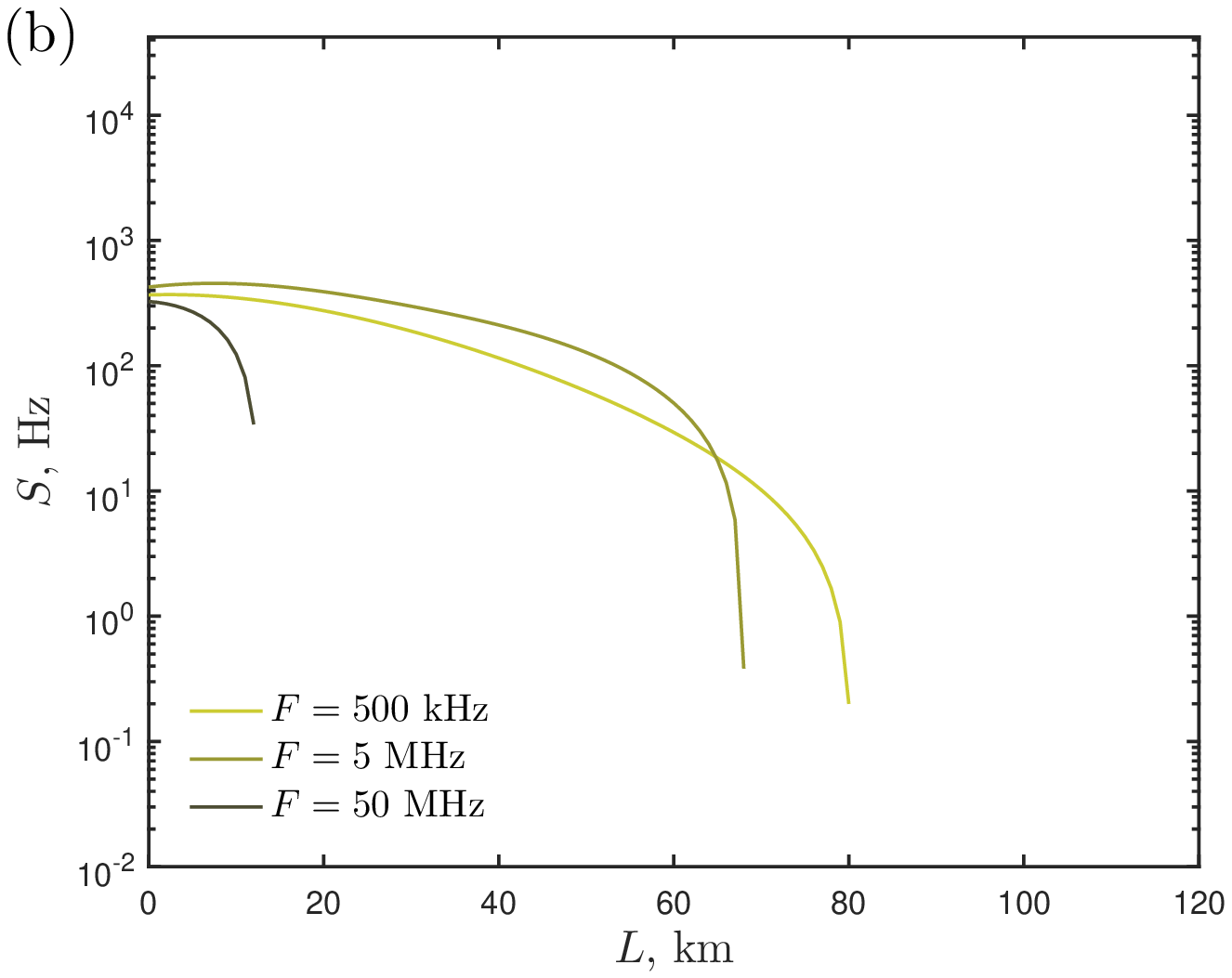}  \\
\includegraphics[scale=0.35]{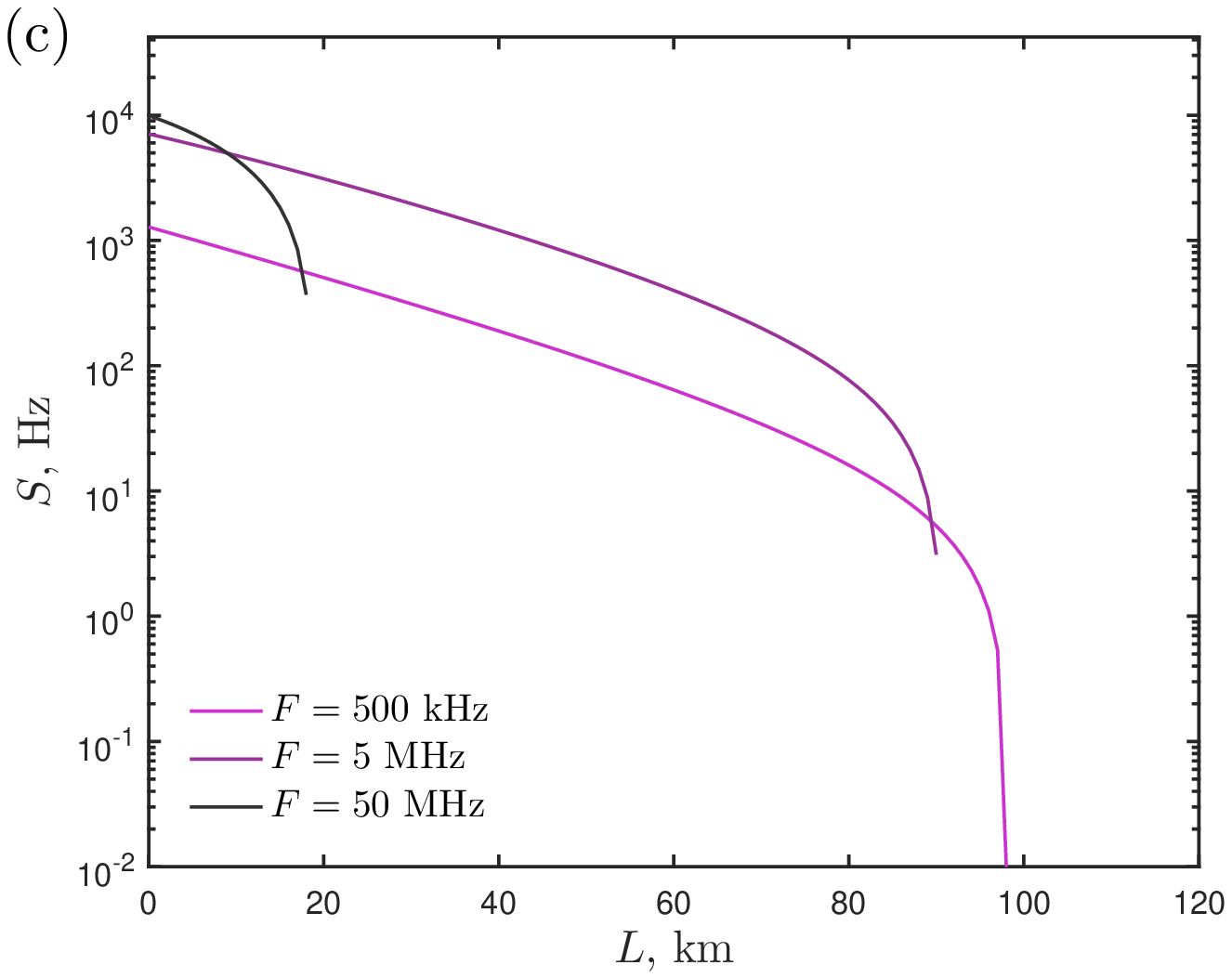}&\includegraphics[scale=0.35]{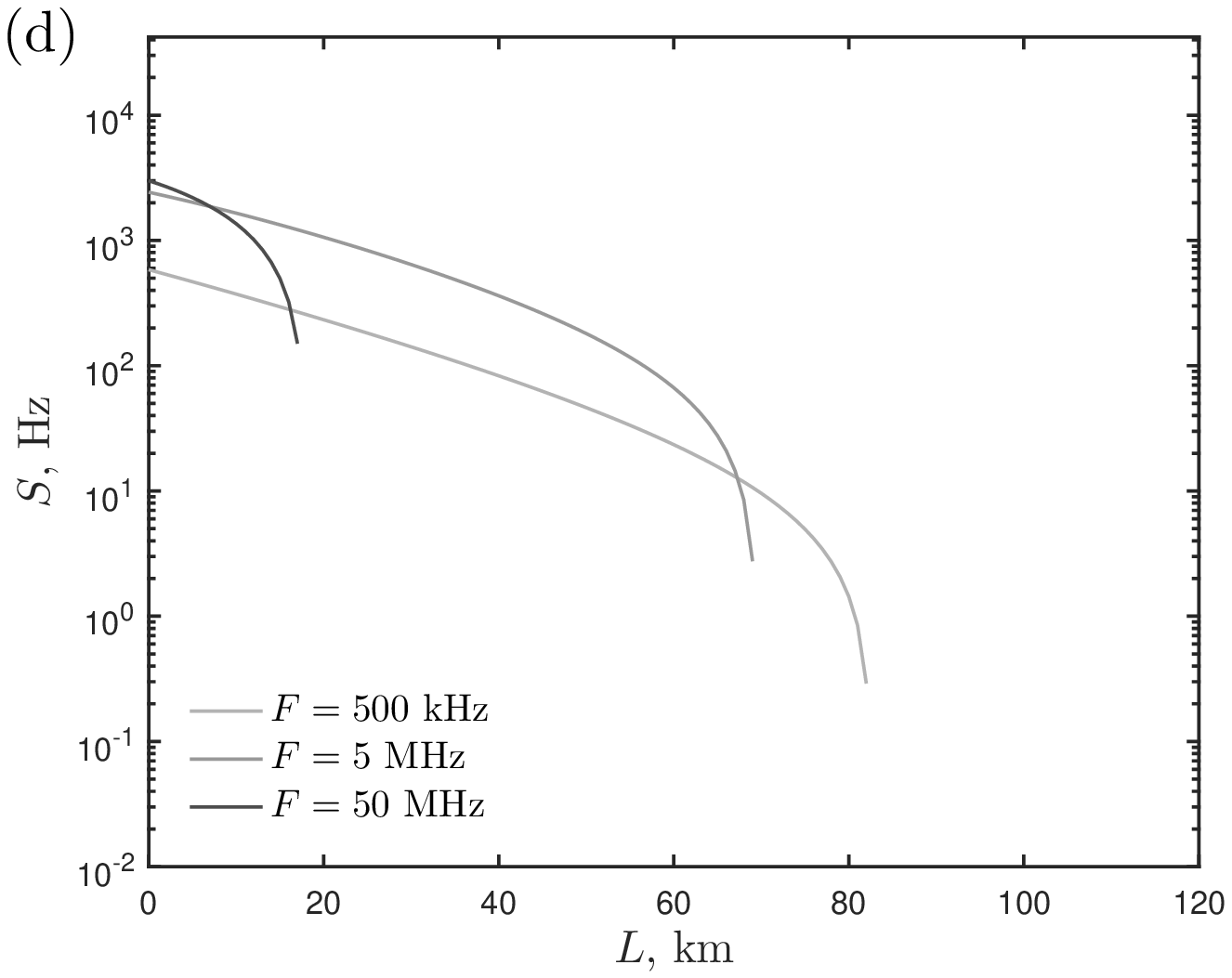}
\end{tabular}
\caption{\small Secure Key rate when no optimization
of the  parameters is performed: (a) BB84, (b) SARG04, (c) decoy BB84, (d) decoy SARG04. Results are reported for three different operating frequencies and as a function of distance.}\label{F:3}
\end{figure}

Before discussing optimization, we present the results for the protocols of interest using typical values for the parameters, {\it{i.e.}} no-optimized. For the internal parameters of the detector, we use the values from Table~\ref{T:1}. We set $T_S=500\,\mu$s, $T_{fr}=1.00$~ms, the transmittance value of Bob's module $T_B=0.5$, and the channel attenuation factor, $\alpha=0.2$~dB/km. The average photon number for the standard BB84 and SARG04 protocols is set to $\mu_1=T_c$ and $\mu_1=2\sqrt{T_c}$ respectively. For the decoy version of these protocols, we use three states with the following parameters: $\mu_1=0.4$ ($\epsilon_1=0.93$), $\mu_2=0.001$ ($\epsilon_2=0.05$) and $\mu_3=0$ ($\epsilon_3=0.02$). Error correction factor value $f=1.1$. The secret key rates where calculated for three different frequencies, $F_1=500$~kHz, $F_2=5$~MHz and $F_3=50$~MHz are reported in Fig.~\ref{F:3}. To get these results we used  $\Delta_t=10\ \mu$s for the BB84 protocols and $\Delta_t=20\ \mu$s for SARG04 protocols, which ensures acceptable secure key rates ($S>0$). Note that, as far as distance is concerned, the higher the operating frequency, the worse performance of the protocol.
   
\begin{figure}[h!]
\centering
\begin{tabular}{cc}
\includegraphics[scale=0.35]{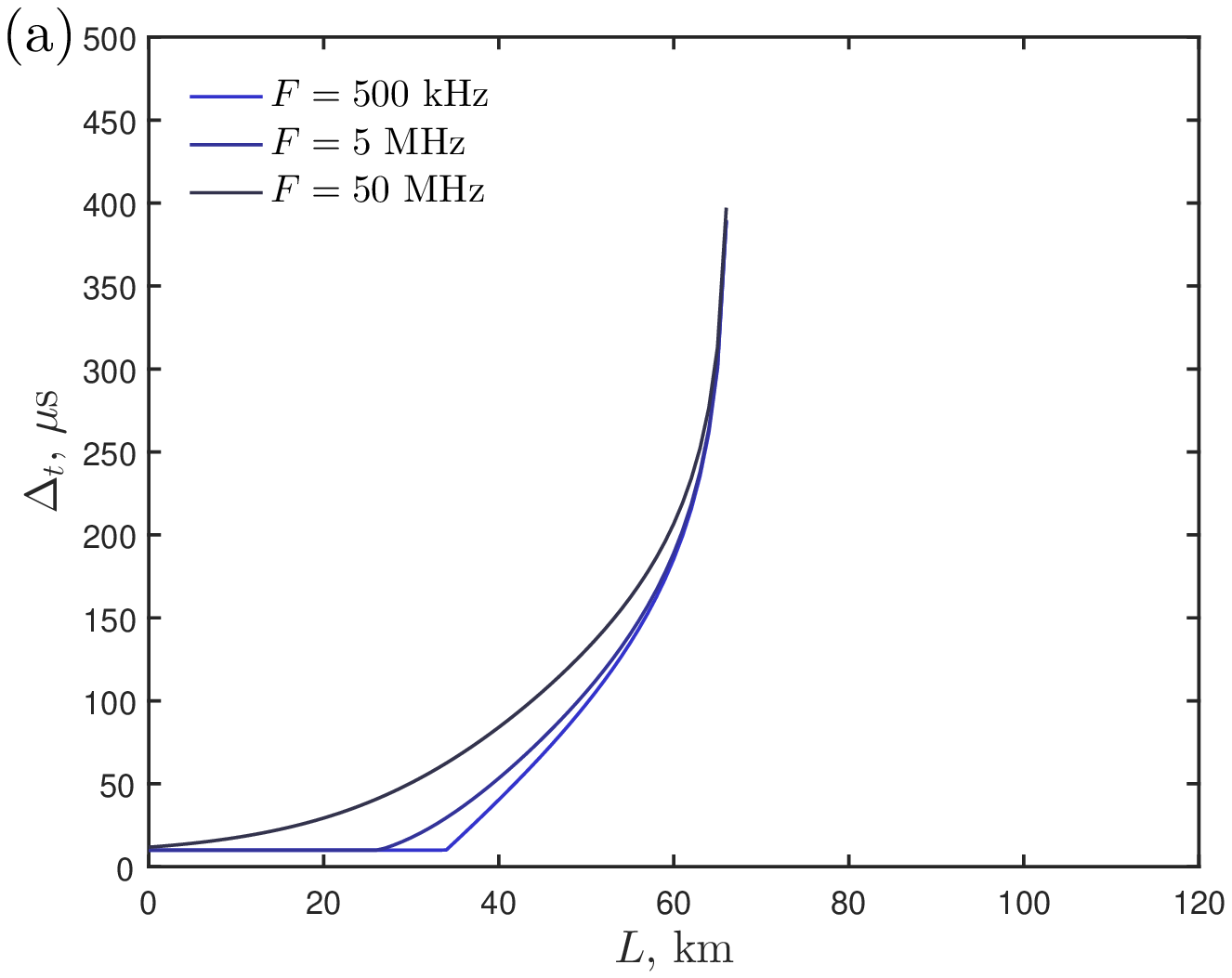}&\includegraphics[scale=0.35]{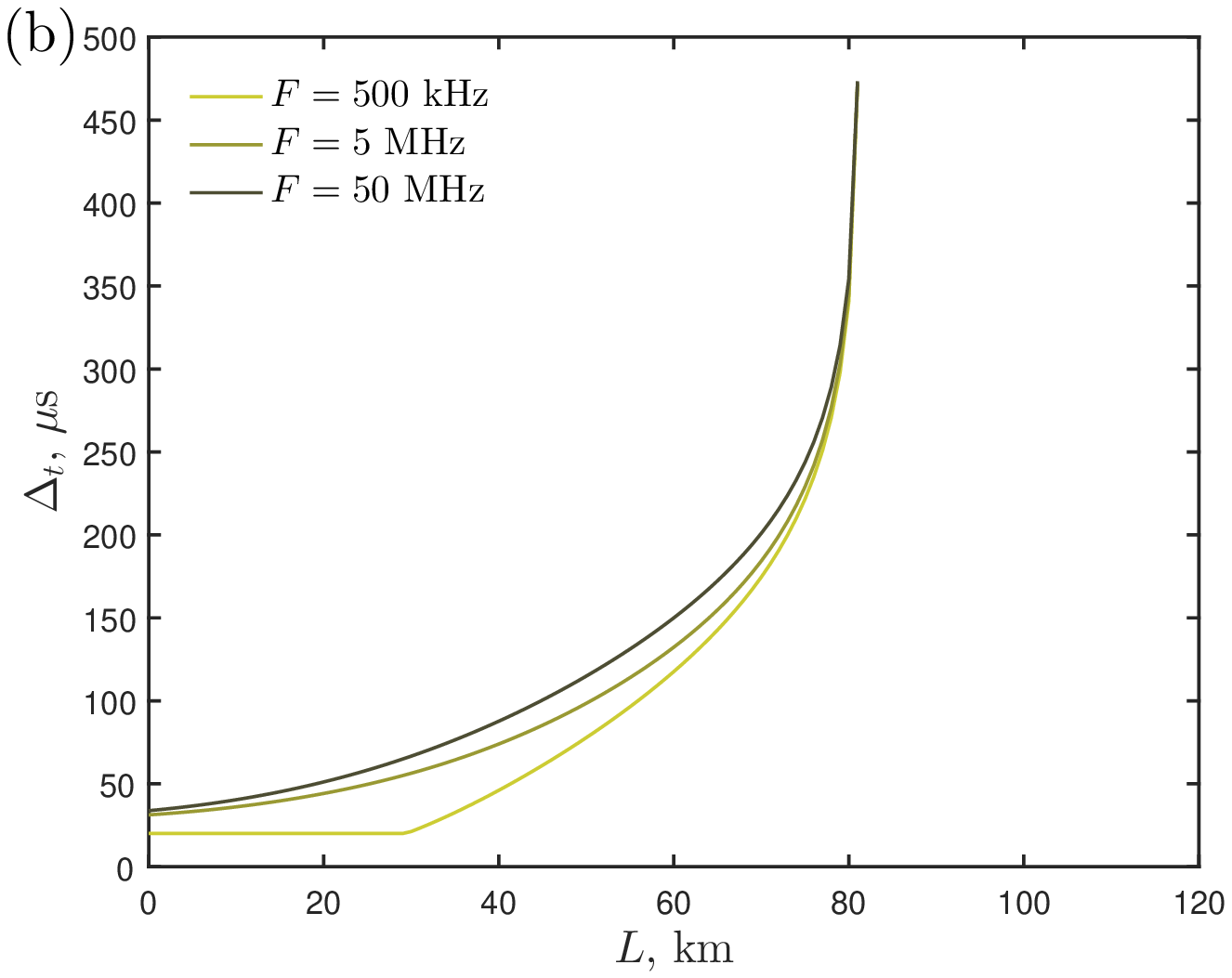}  
\end{tabular}
\caption{\small Optimal value of the dead-time ($\Delta_t$) for the standard version of the (a) BB84, (b) SARG04 protocols as a function of distance and for three different operating frequencies. }\label{F:8}
\end{figure}

\begin{figure}[h!]
\centering
\begin{tabular}{cc}
\includegraphics[scale=0.35]{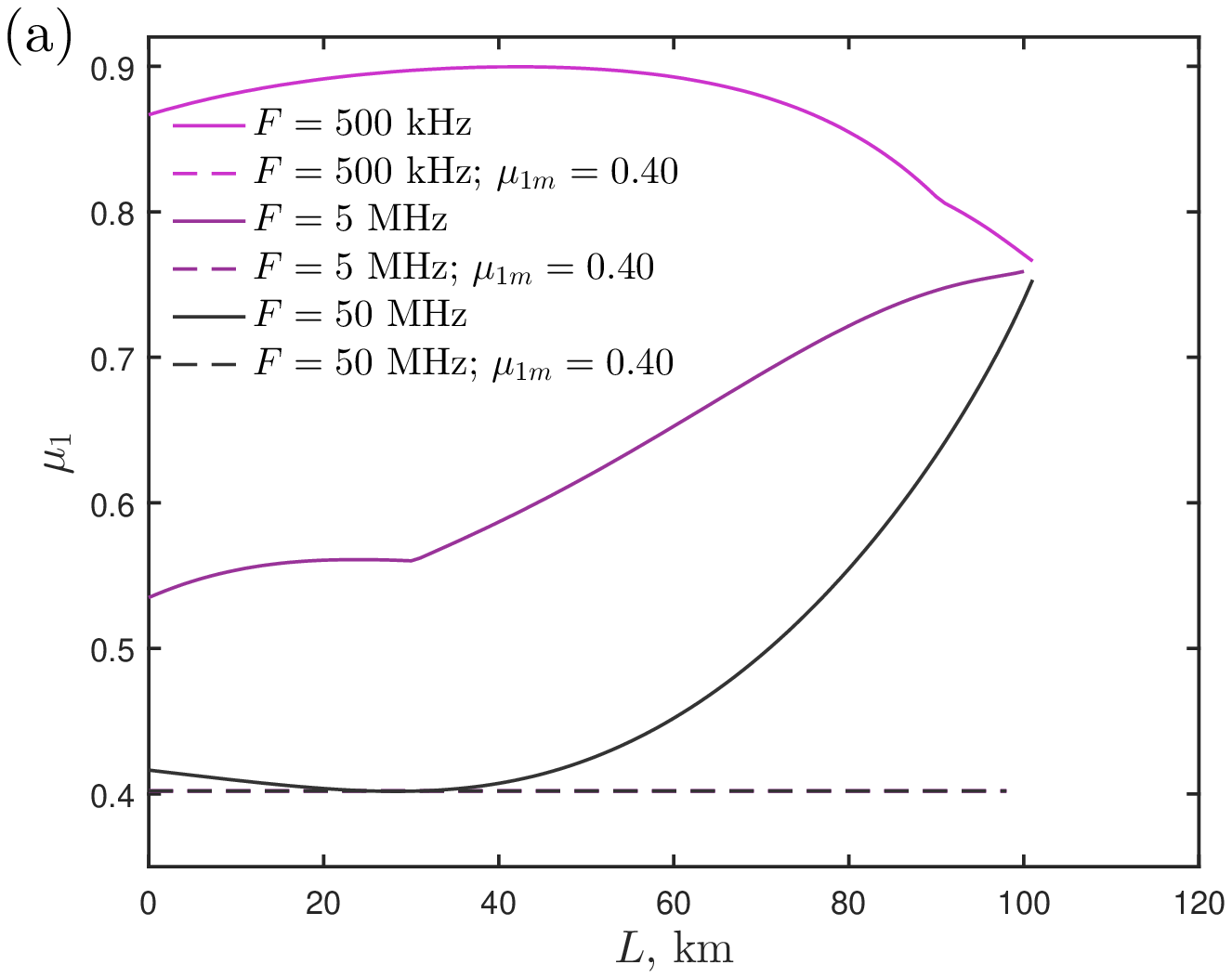}&\includegraphics[scale=0.35]{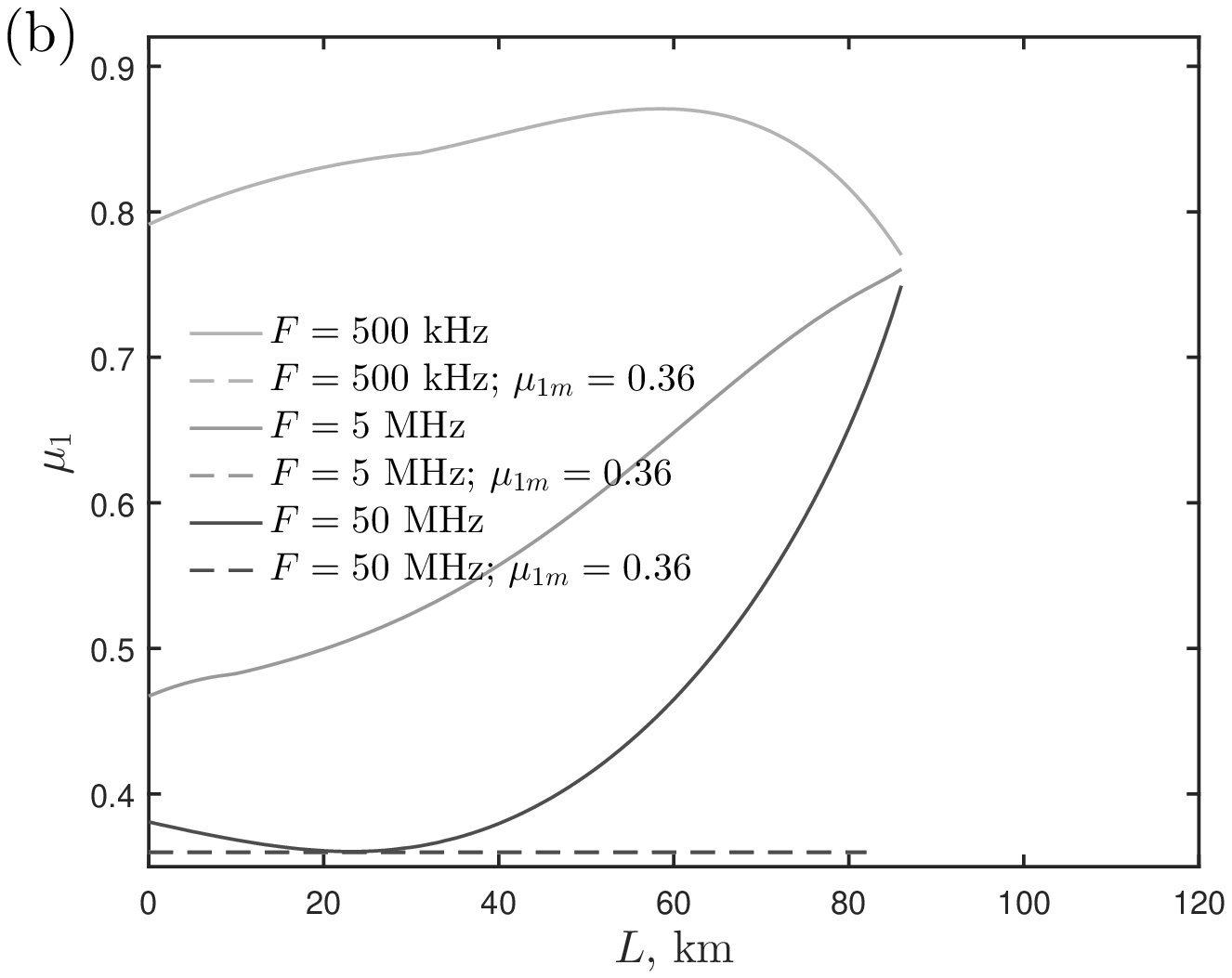}\\
\includegraphics[scale=0.35]{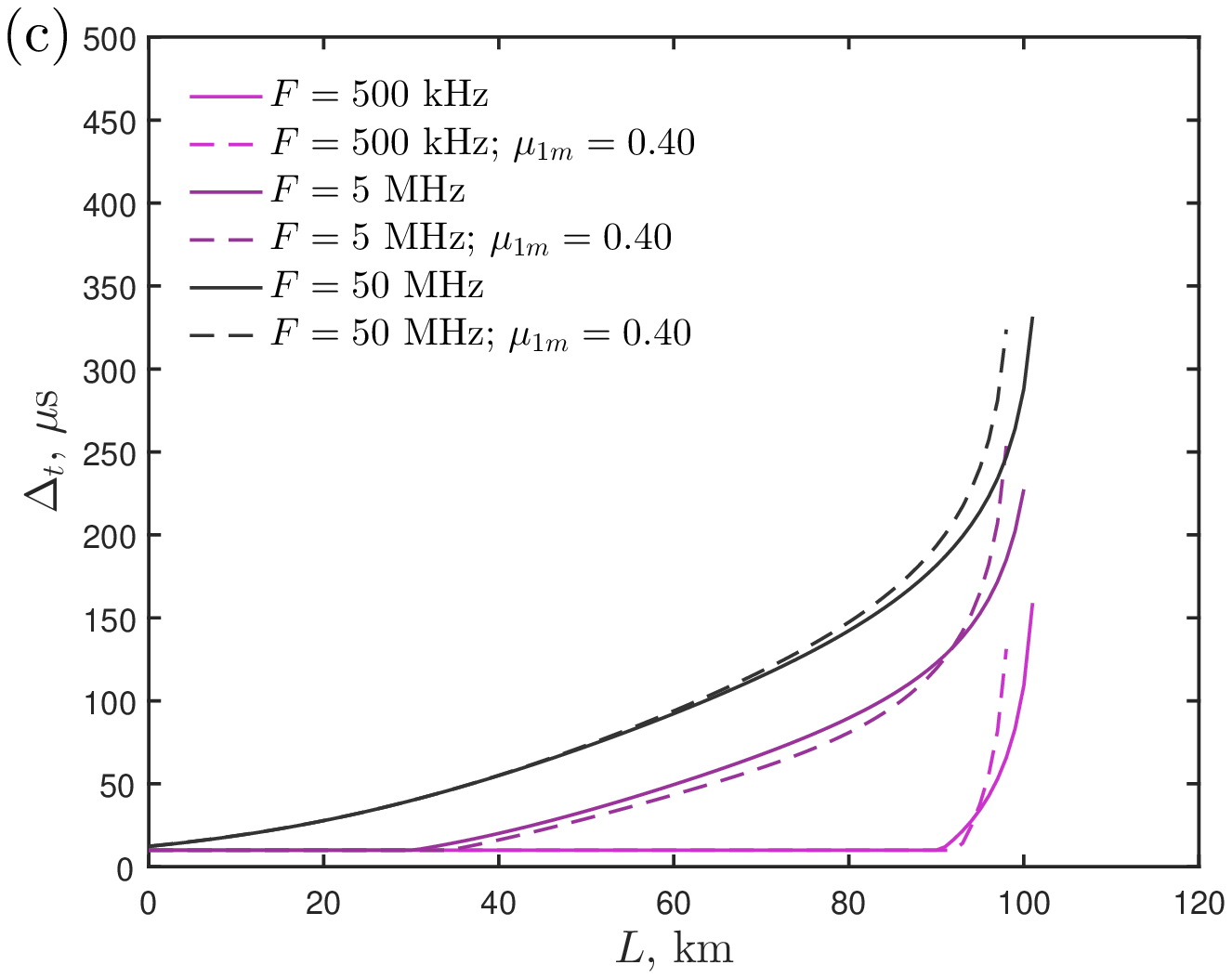}&\includegraphics[scale=0.35]{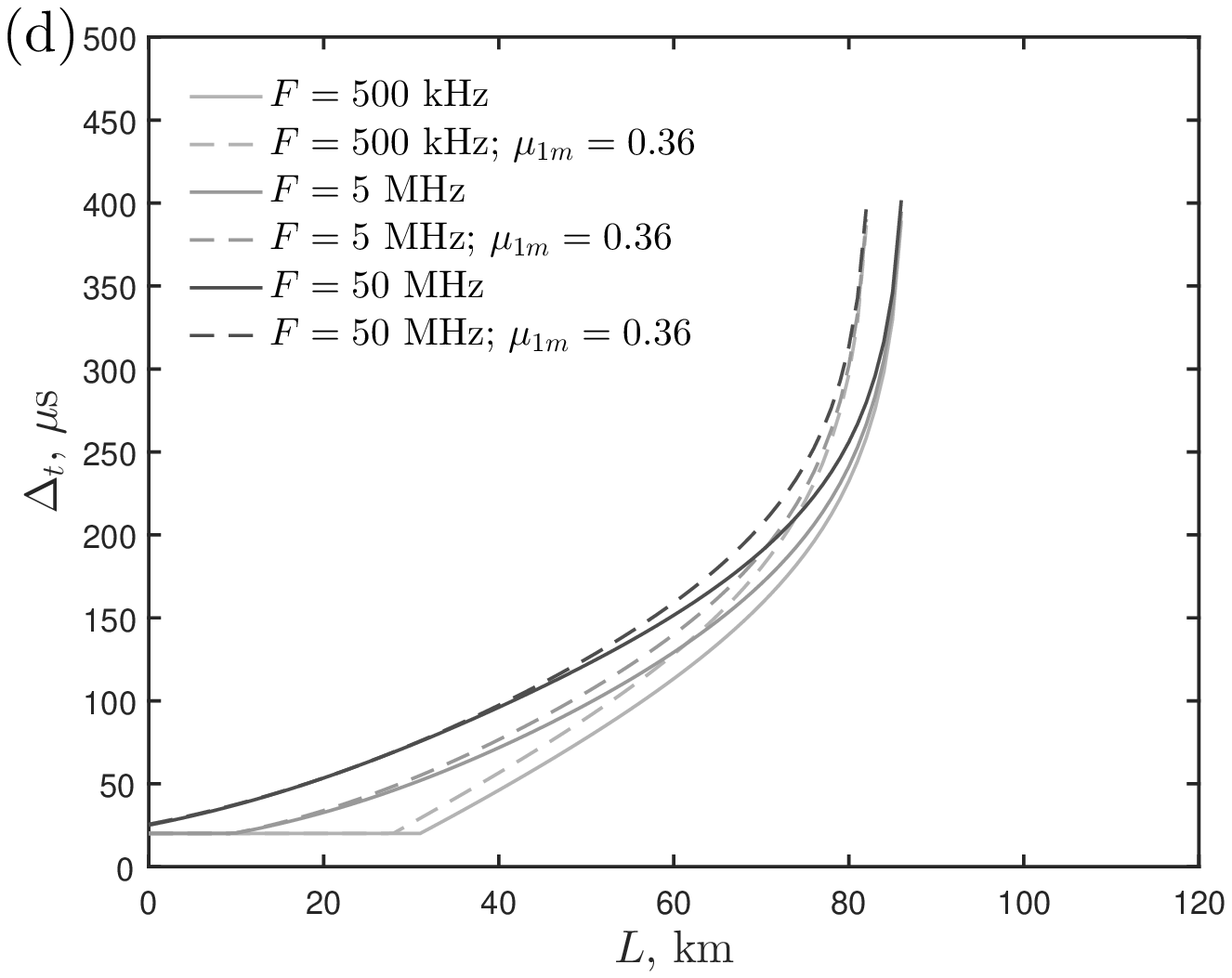}
\end{tabular}
\caption{\small Optimal value of the mean photon-number ($\mu_1$) as a function of distance and for three different operating frequencies: (a) decoy BB84, (b) decoy SARG04. Optimal value of the dead-time ($\Delta_t$) as a function of distance and for three different operating frequencies: (c) decoy BB84, (d) decoy SARG04.The dashed-line curves are obtained using the minimal optimal value of the mean photon-number $\mu_{1m}$, which is kept fixed regardless the distance, and operating frequency value. }\label{F:9}
\end{figure}

Let us now turn to optimization, which relies on maximizing the secure key rate $S$. To this end, we optimize $\Delta_t$ in the standard version of the BB84 and SARG04 protocols, and both  $\mu_1$ and  $\Delta_t$ int their decoy versions. Results for the optimal values of $\Delta_t$ and $\mu_1$, as a function of distance  are reported in Figs.~(\ref{F:8},\,\ref{F:9}) respectively. 
In the optimization process, the dead-time is the most relevant parameter to minimize the afterpulsing effect in the optimal distance. In the end, the maximum distance range is limited mainly by channel losses and dark counts probability of the detectors.

\begin{figure}[h!]
\centering
\begin{tabular}{cc}
\includegraphics[scale=0.35]{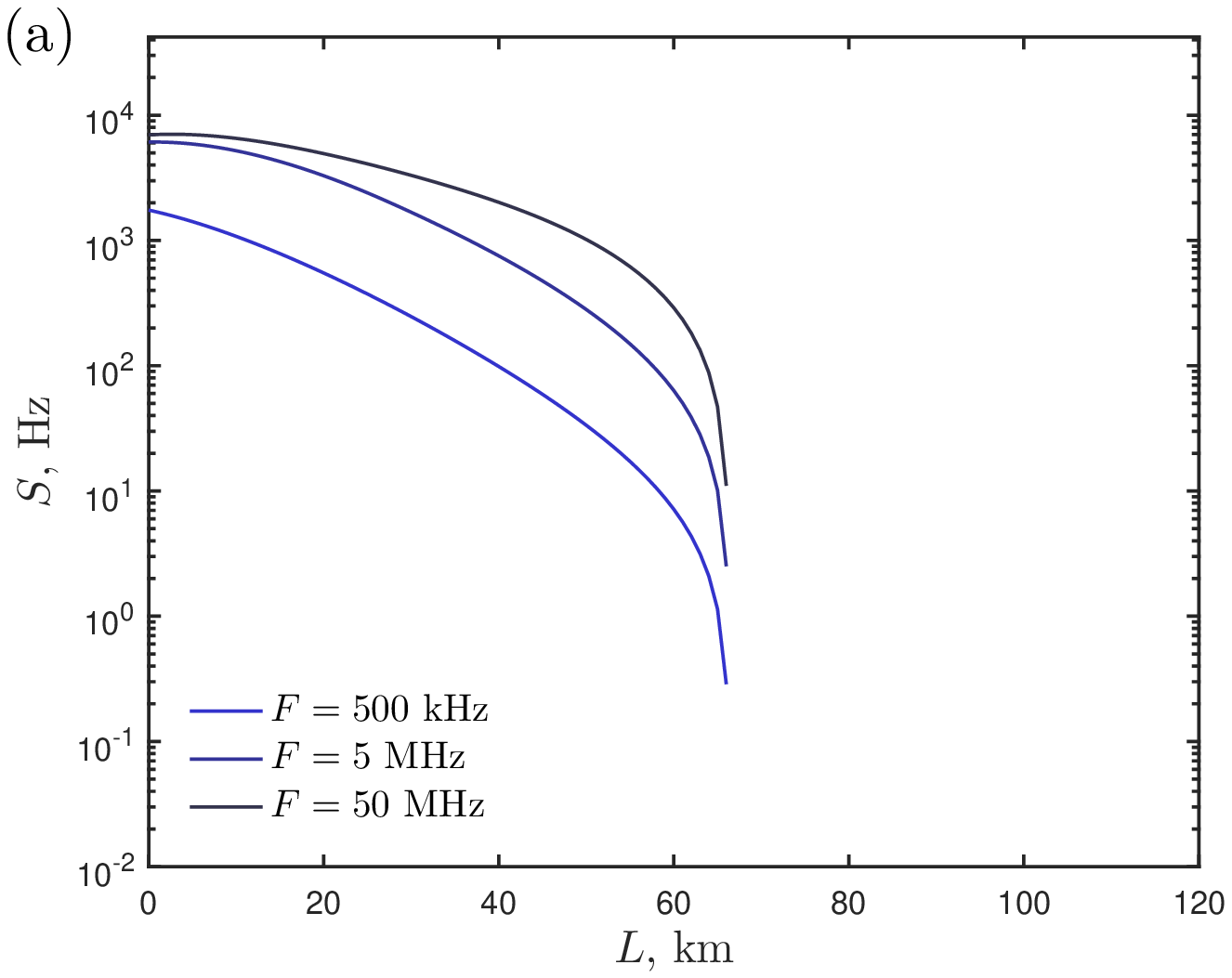}&\includegraphics[scale=0.35]{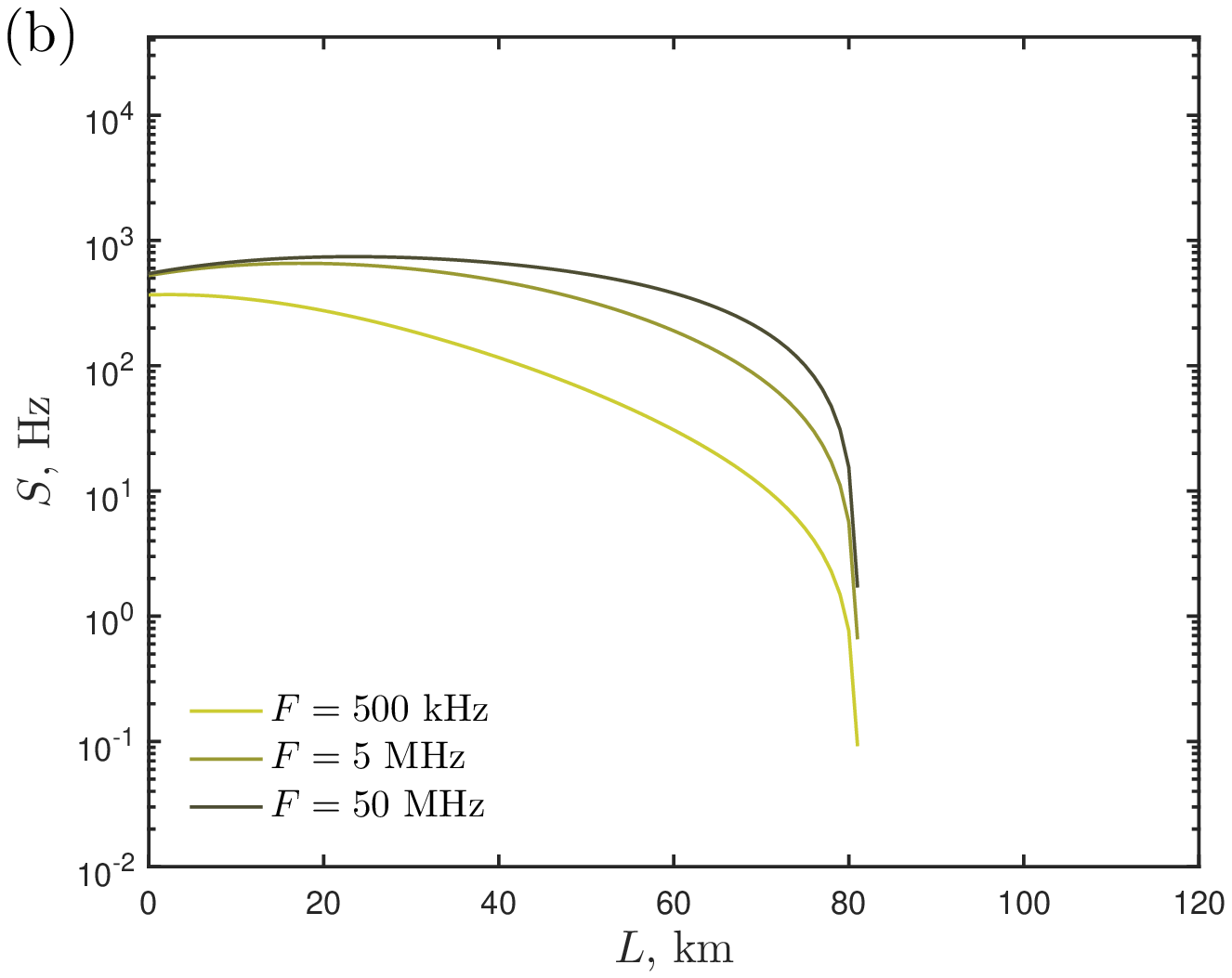}  \\
\includegraphics[scale=0.35]{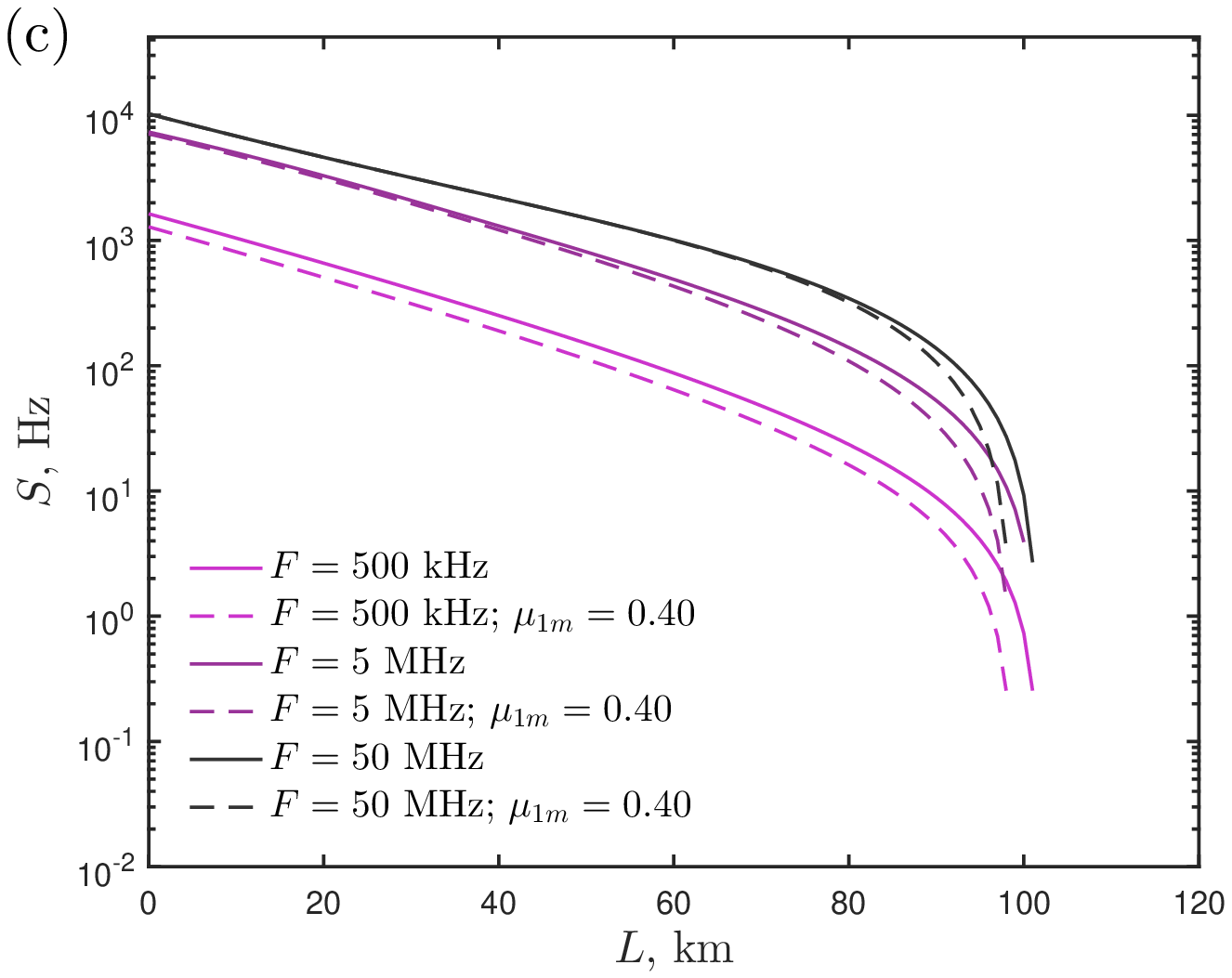}&\includegraphics[scale=0.35]{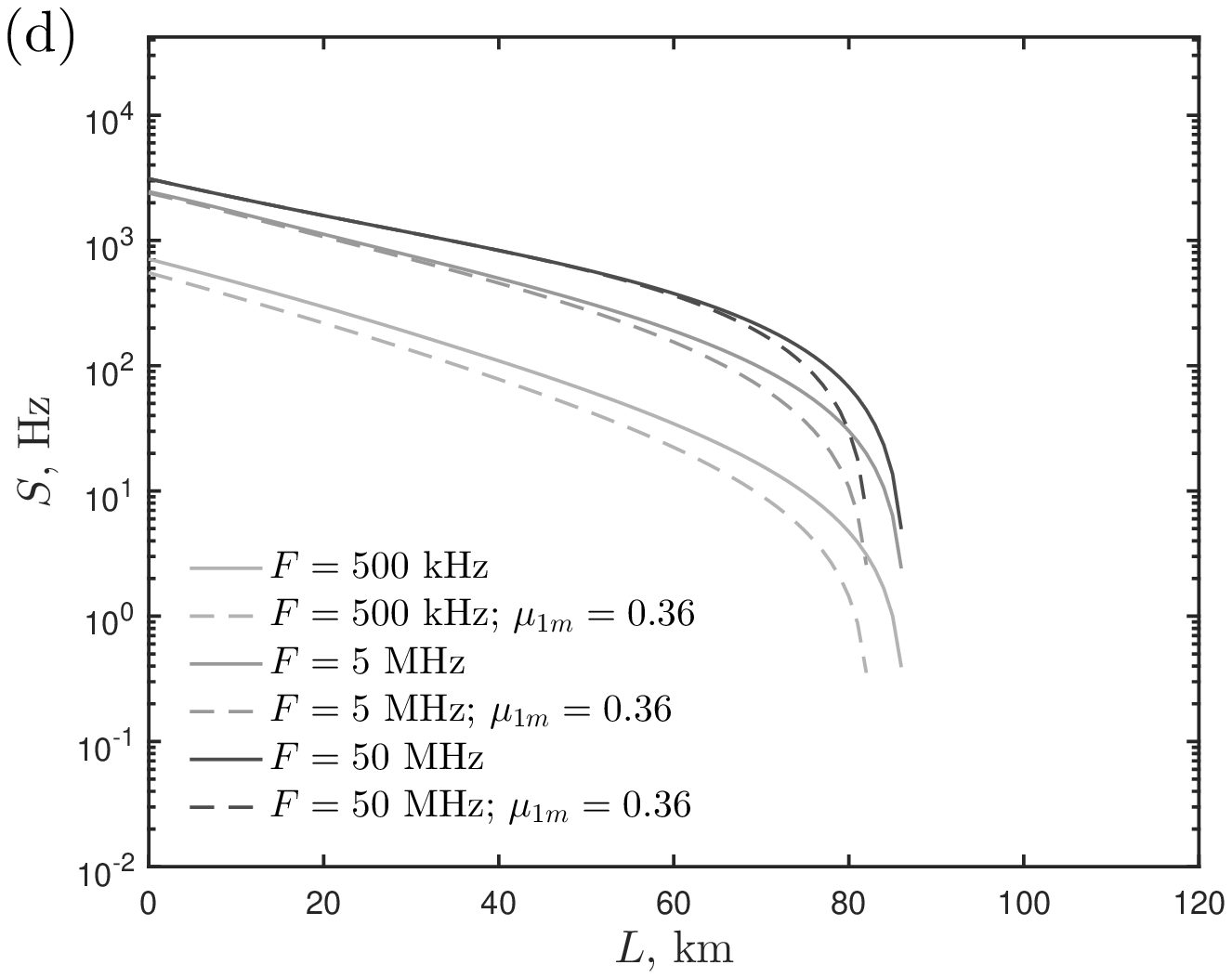}
\end{tabular}
\caption{\small Optimized secure key rate ($S$) as a function of distance for different values of the operating frequency: (a) BB84, (b) SARG04, (c) decoy BB84, (d) decoy SARG04. The dashed-line curves are obtained with the value of the mean  photon number fixed to its minimum value $\mu_{1m}$, independent of the distance and operating frequency value.}\label{F:4}
\end{figure}

\noindent The results for the secret key rate when the optimization is implemented are shown in Fig.~\ref{F:4}. The reported results correspond to the same frequencies and the same parameters as those used in Fig.~\ref{F:3}, except for  $\Delta_t$ and  $\mu_1$. For the dead-time, the optimal values that are used are reported in Figs.~\ref{F:8} and~\ref{F:9}, for each distance. For the average number of photons we use $\mu_1=T_c$ and $\mu_1=2 \sqrt {T_c}$, in the standard BB84 and SARG04 protocols, respectively. In the decoy version of the protocols, the optimal values of the mean photon-number (solid lines) are reported in Fig.(\ref{F:9}). We performed the optimization only of the dead-time value using the minimum value of the mean-photon number ($\mu_{1m}$), regardless the distance or operation frequency values. These results of optimization and secret key rates are presented using dashed lines in Figs.~\ref{F:9} and~\ref{F:4}. We chose keeping fixed the minimal value to prevent the PNS attack. 

\begin{figure}[h!]
\centering
\begin{tabular}{cc}
\includegraphics[scale=0.35]{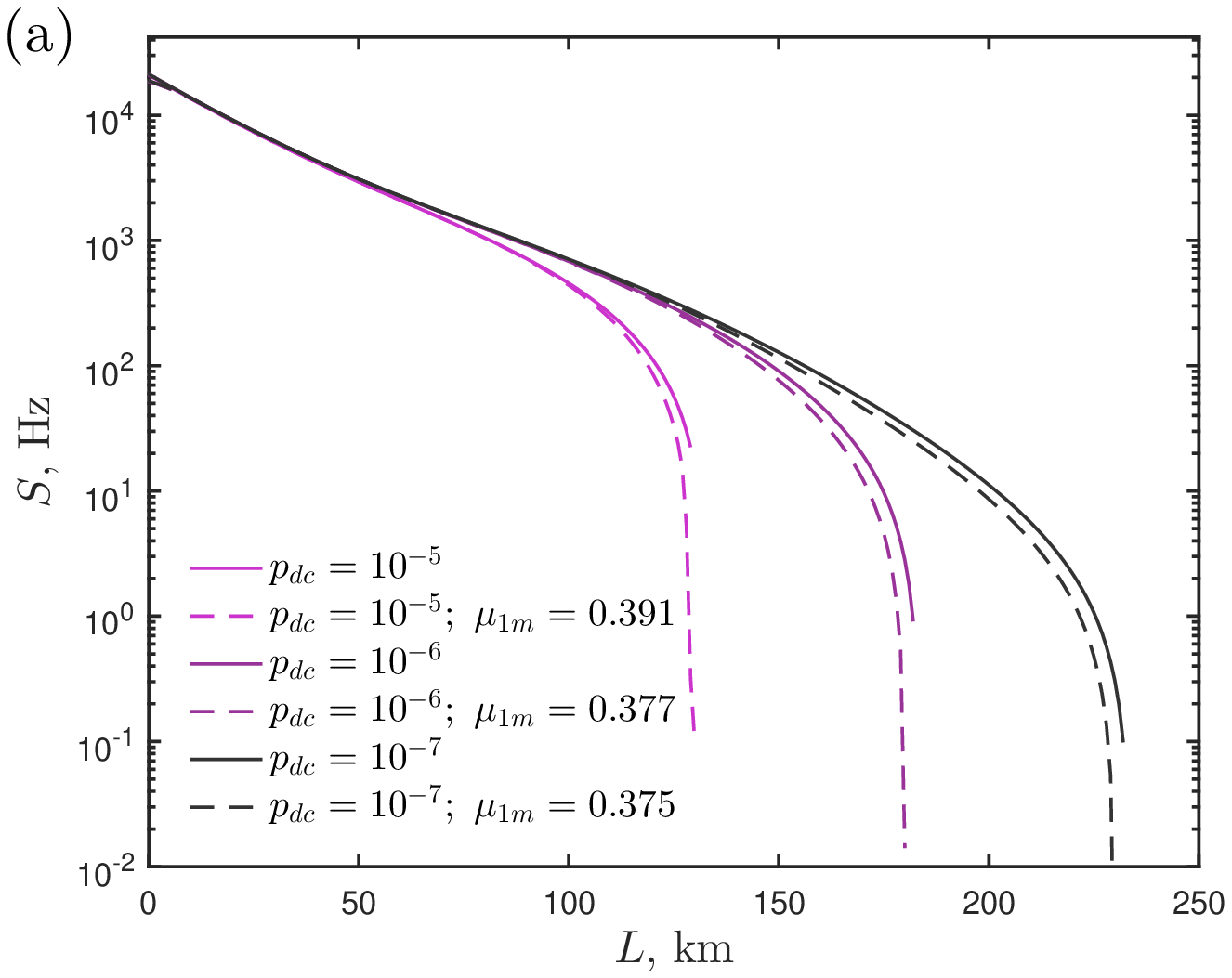}&\includegraphics[scale=0.35]{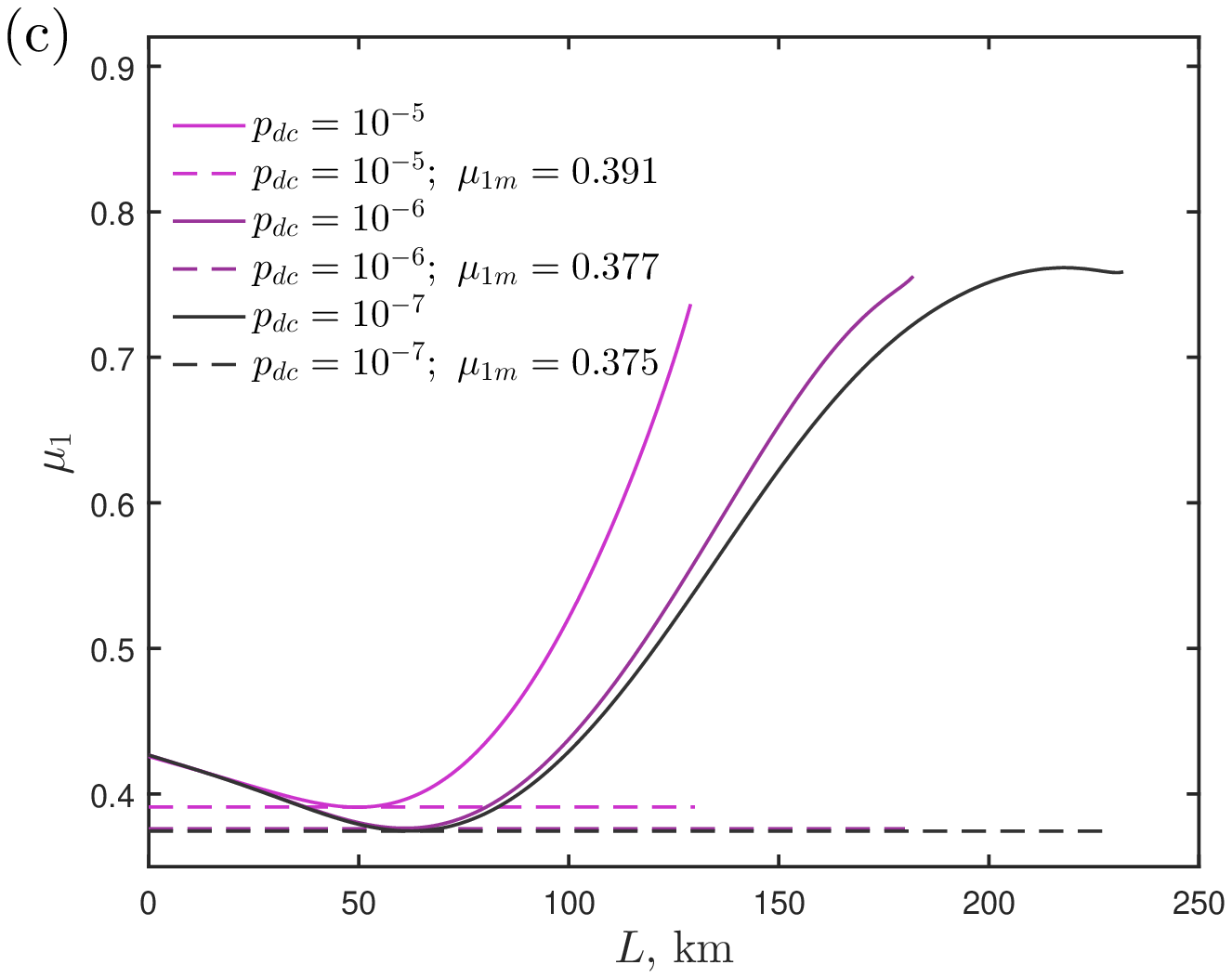}  \\
\includegraphics[scale=0.35]{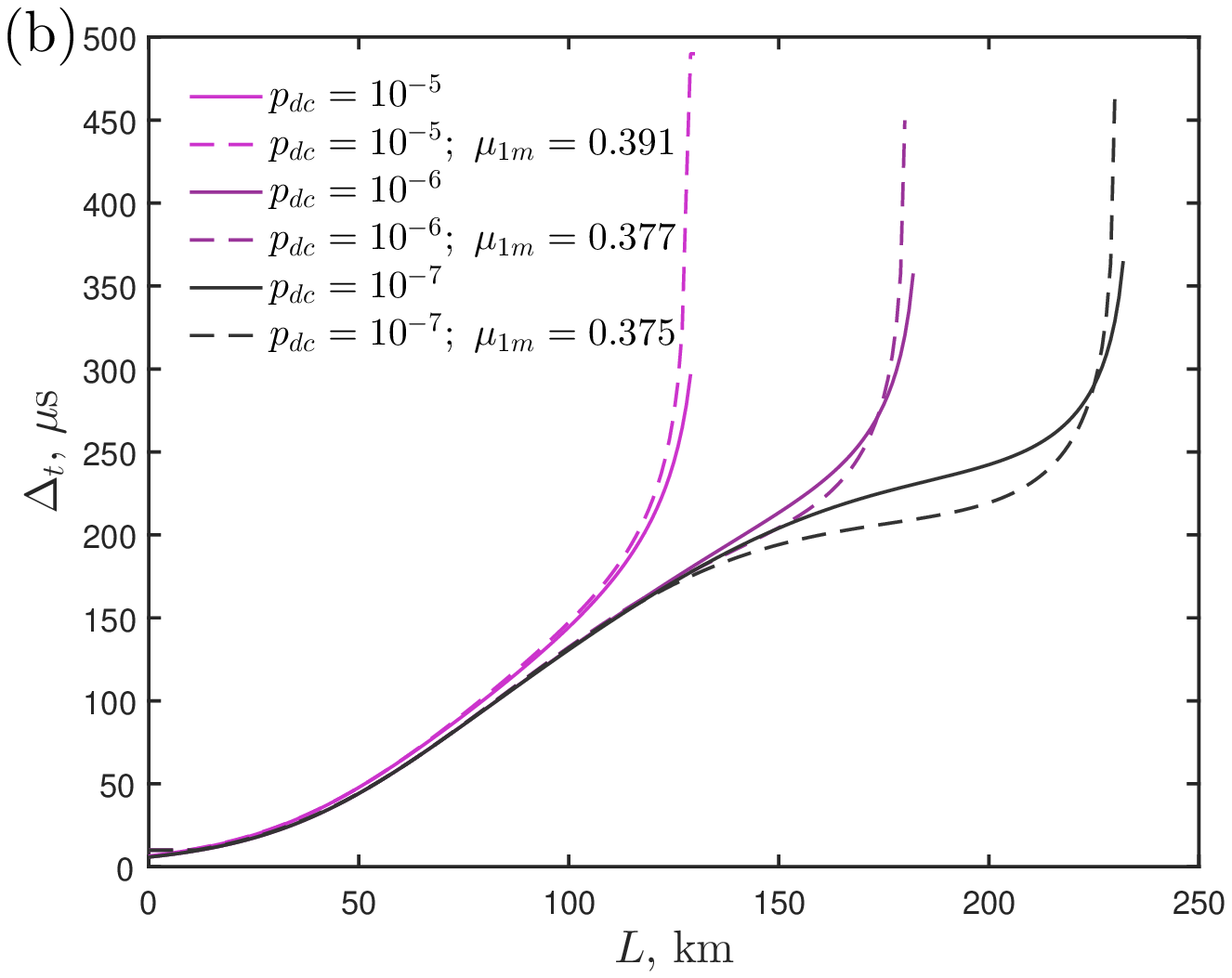}&\includegraphics[scale=0.35]{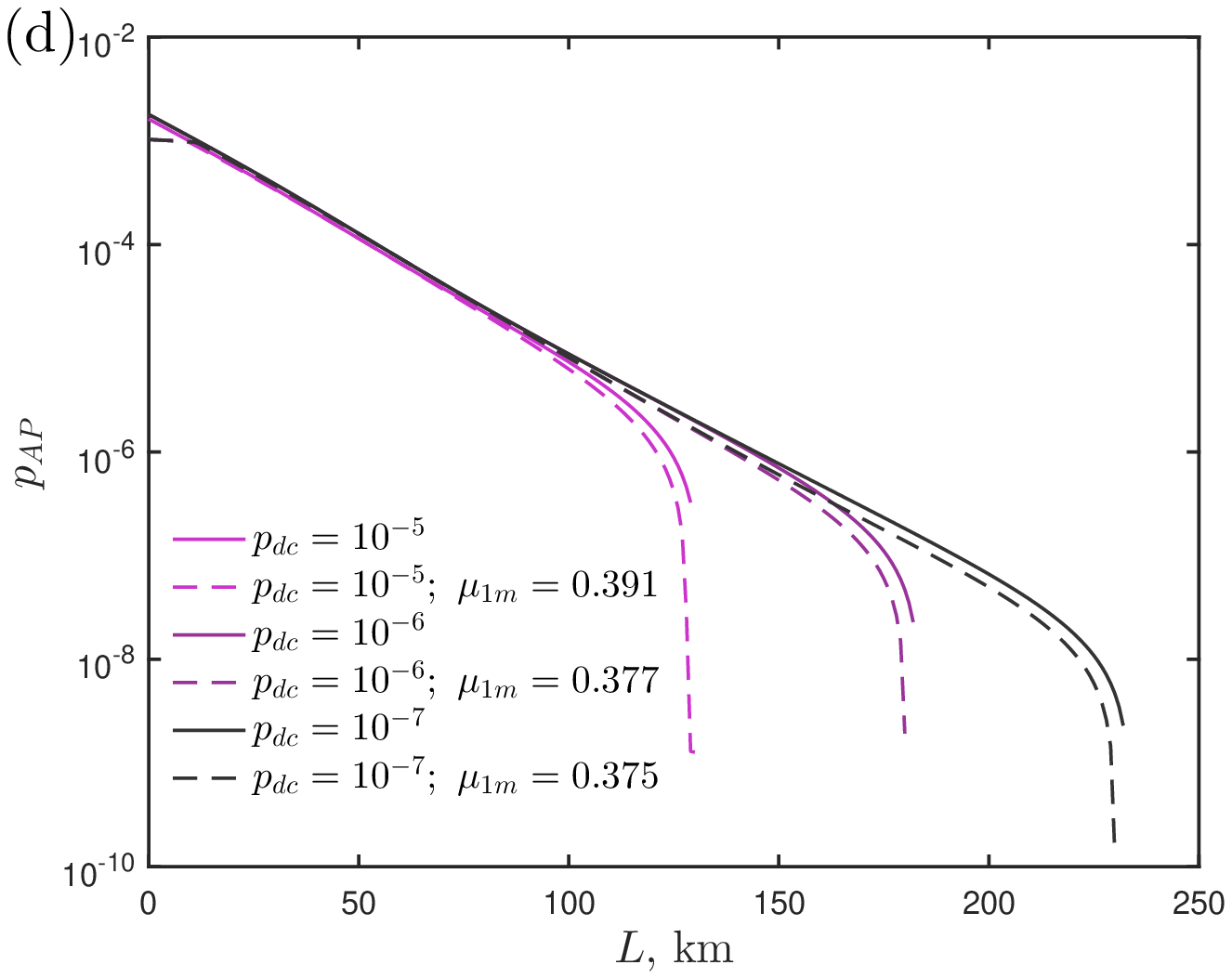}
\end{tabular}
\caption{\small Optimized secure key rate ($S$) as a function of distance for different values of the operating frequency: (a) secure key rate ($S$), (b) dead-time ($\Delta_t$), (c) mean photon-number ($\mu_1$), (d) afterpulsing probability ($p_{AP}$). The dashed-line curves are obtained with the value of the mean  photon number fixed to its minimum value $\mu_{1m}$, independent of the distance, but it depends on dark-count probability.}\label{F:AA}
\end{figure}

To show that optimizing dead-time value to maximize the secret key rate, minimize the impact of the afterpulsing in the secure distance range. We present the results of the numerical optimization of dead-time and mean photon-number values of the performance of the three state decoy BB84 protocol, using the same operation parameters beside $F=50$~MHz, $\eta=0.2$, and using three different values of dark count probabilities $p_{dc}\in \{10^{-5},\ 10^{-6},\ 10^{-7}\}$. The results of the secret key rate $S$ are presented in Fig.~\ref{F:AA}(a). The optimal dead-time and mean photon-number values are shown in Figs.~\ref{F:AA}(b) and~\ref{F:AA}(b), respectively. Finally the residual afterpulsing probability after the dead-time optimization is depicted in Fig.~\ref{F:AA}(c)

\subsection{Monte-Carlo simulation vs model }
\begin{figure}[h!]
\centering
\begin{tabular}{c}
\includegraphics[scale=0.35]{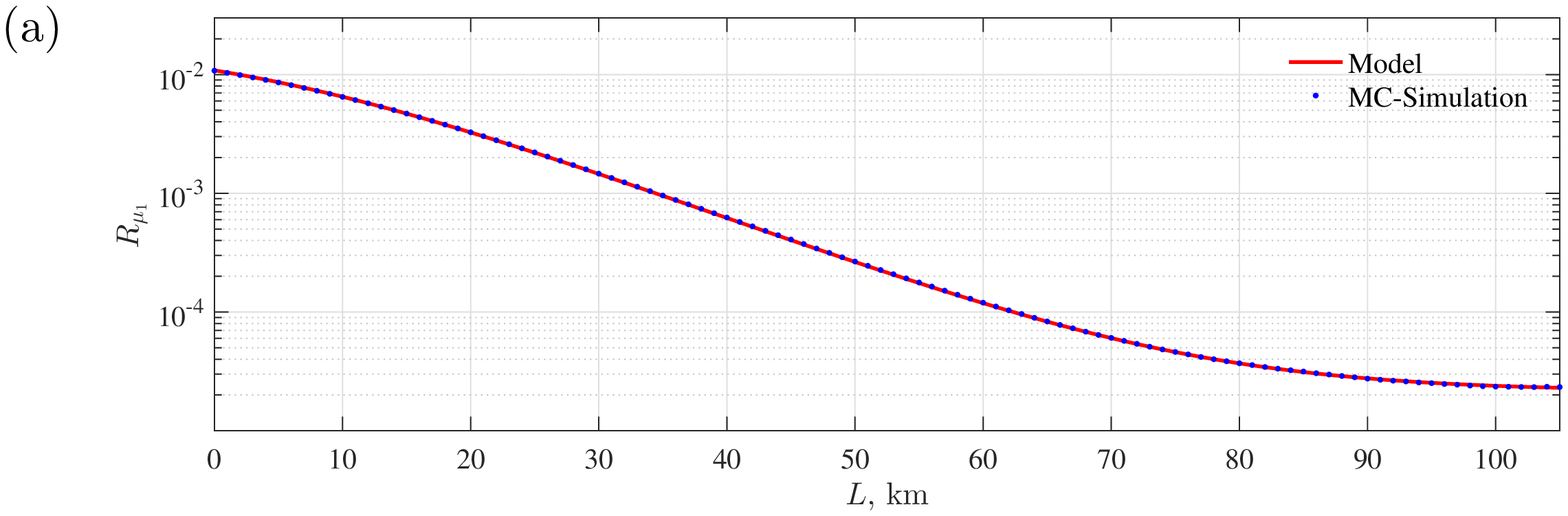}\\
\includegraphics[scale=0.35]{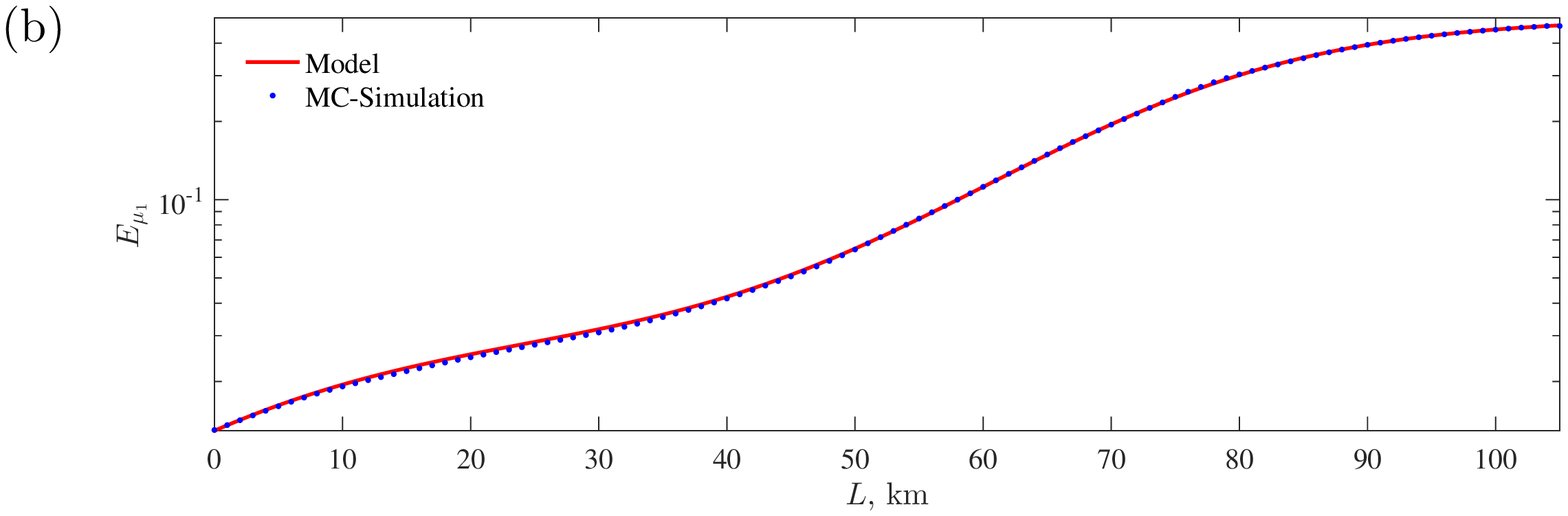}
\end{tabular}
\caption{\small Model (solid red lines) vs MC simulation  (dashed blue lines) results for a standard BB84 protocol: (a) sifted key rate ($R_{\mu_1}$) with deviation between model and MC simulation $\sigma_e=0.011$; and (b) total QBER ($E_{\mu_1}$) with relative deviation between model and MC simulation $\sigma_e=0.013$. }\label{F:1}
\end{figure}

For completeness, in this section we compare our results, based on the model described above with the Monte Carlo (MC) simulations of the P\&M-QKD protocols. We start by specifying the values of the parameters required for the simulation and indicating their origin. Then we describe the algorithm and report the results in Fig.~\ref{F:1}. 

Inputs parameters:
\begin{itemize}
    \item Parameters inherent to the detector: $\eta$, $p_{dc}$, $Q$ and $\tau$. 
We use the values reported in Table~\ref{T:1}.
    \item Quantum channel parameters: $\alpha$, $L$, we use values of silica optical fibers $\alpha=0.2$~dB/km and variable length $L\in\{0,\ 120\}$~km.
    \item Timing parameters: $F= 5$~MHz, $\Delta_t = 10\ \mu$s, $T_S=500\ \mu$s, $T_{fr}=1.00$~ms.
    \item Sources parameters for BB84 protocol, mean photon number $\mu_1=T_c=10^{-\alpha L/10}$  and occurrence probability $\epsilon_1=1$.
    \item Phases $\phi_i^{A}\in\{0,\ \pi/2,\ \pi,\ 3\pi/2 \}$ and $\phi_j^{B}\in\{0,\ \pi/2 \}$ and occurrence probabilities $\theta_i^{(A)}=1/4$, $\theta_j^{(B)}=1/2$, for different values of $i$ and $j$.
    \item Bob's module transmittance $T_B=0.5$, and error correction factor $f=1.1$.
    \item Number of frames $N_{fr}= Ft_fr$.
    \item Number of gates in the frame: $N_S=T_SF$ and number of gates in the dead-time $N_d=\Delta_tF$.
\end{itemize}

Besides the input listed in previous lines, the following steps describe the implementation of the MC simulation:
\begin{enumerate}
\item Use the random number generator and the probability distributions for $\epsilon_k$, $\theta_i^{(A)}$ and $\theta_j^{(B)}$ to create the state configuration $\mu_k$, $\phi_i^{(A)}$ and $\phi_j^{(B)}$. The probability of detection per gate for this state  is given by $p(m)\leftarrow 1-\exp\left(-\gamma_{ijw}\mu_k\right)$.
\item Add dark count probability: $p(m)\leftarrow 1-(1-p_{dc})(1-p(m))$.
\item  For  a given gate $m$, determine if an event occurs by comparing $p(m)<rand()$, a randomly generated number.
\item If an event occurs: add 1 to the detector counter in the $m$-position of the frame, jump $(N_d+1)$-gates and add afterpulsing contribution $p(m')\leftarrow 1-(1-p(m'))(1-Q/\tau\exp(-(m'-m)/(F\tau)))$, for all $m'\in \left[N_d+n,\ N_S\right]$. 
\item If no event occurs, increase $m\leftarrow m+1$, return to step 1 until all gates in the frame are evaluated $m=N_{S}$.
\item Sifted key. Discard events according to each P\&M-QKD protocol and store the surviving events. 
\item Add all detection events in the frame and dividing by  $N_S$, to obtain $R_{\mu_1}$.
\item QBER: {\it{i}}) count the number of bits in the sifted keys  that are correct and divide the total by $N_S$ to get $N_{\mu_1}$, {\it{ii}}) evaluate $E_{\mu_1}=1-N_{\mu_1}/R_{\mu_1}$
\item Repeat the steps 1 to 8 $N_{fr}$-times, and average the values obtained for $R_{\mu_1}$ and $E_{\mu_1}$.
\end{enumerate}

\noindent  MC simulation is a robust method to evaluate the performance of QKD protocols, but one of its drawbacks is that it takes too much time to perform a systematic, complete analysis of running data. However this tool is undoubtedly useful as a benchmark for other treatments or approximate calculations. It is in the sense that we compare the results of our model with those obtained from the MC simulation for the performance of the standard BB84 protocol.The results for the sifted key rate, $R_{\mu_1}$, and the total QBER, $E_{\mu_1}$, are shown in Fig.~\ref{F:1}. Identical operating conditions (values of the parameters) were used in the MC simulation (dashed blue lines) and the model (solid red lines) to obtain these curves. The results agree over the entire range of distances with a relative deviation of the parameter $\sigma_e=0.011$ for $R_{\mu_1}$ and $\sigma_e=0.012$ for $E_{\mu_1}$, defined as,
\begin{equation}
    \sigma_e=\sqrt{\frac{1}{M-1}\sum_{n=1}^M\left[ \frac{X_n^{(sim)}-X_n^{(mod)}}{X_i^{(mod)}} \right]^2}
\end{equation}
where $X_n^{(sim)}$ is the value of the $n$-simulated parameter and $X_n^{(mod)}$ is the value of the $n$-calculated parameter using our model.

\section{Summary and Conclusions} \label{Sec:5}
In summary, in this paper we present a formalism that allows us to incorporate afterpulsing  and dead-time effects in the performance analysis of P\&M-QKD protocols. The analytical model is used to determine the parameters that optimize the performance of the secret key rate. A comparative study of the performance of P\&M-QKD protocols using their optimal parameters is presented, we apply the treatment to analyze the standard BB84, and SARG04 protocols, as well as their decoy-state versions.  We also report the comparison of the results obtained using the formalism developed here with those of a Monte Carlo simulation.

From the analysis carried out, we conclude that:
\begin{itemize}
\item We report an analytical function of the afterpulsing probability in terms of dead-time and timing parameter.
\item The dead-time and the mean photon-number are the  parameters relevant for the optimization process, and the optimization of both parameters should be performed simultaneously.
\item Compared to the case where a fixed dead-time is considered, the remainder of afterpulsing has the effect on the QBER and the secure key generation rate. 
\item The higher the operating frequency the greater the effects of afterpulsing.
\item Increasing the distance range over which protocols are secure requires longer dead-time values. 
\item After this optimization procedure, the secure distance is limited mainly by the channel losses and dark counts probability.
\item The formalism we propose here is a practical and reliable tool for evaluating the effects of afterpulsing and optimizing the secret key rate over the full range of secure communication.
\item The model we develop in this works performs well for $\mathrm{max}\{\gamma_{ijk}\mu_k\}<1$.
\end{itemize}
As a final comment, we mention that afterpulsing quantification can be used as a parameter to detect eavesdroppers' activities.  Although eavesdropping strategies involving total blinding attacks do not produce afterpulsing by themselves~\cite{Lyd1,Lyd:10}, other strategies can modify the contribution of afterpulsing~\cite{Wie:1,Jai11,Jai:1}.


\begin{appendix}
\section{Average afterpulsing contribution}\label{Ap:A}
In this Appendix, we provide some details regarding to the derivation of the average contribution of afterpulsing. In order to simplify the notation, the $w$-label is removed from the SPDs identifier. Below we consider one single frame, for which the afterpulsing probability is described by a discretized single exponential decay function,
\begin{equation}
p_{af,m}=kQ\exp\left( -mkt_F\right), \label{Eq:Af}
\end{equation} 
where, $k=\tau^{-1}$. After a detection event, this probability decreases with the number of gates ($m$) with period $t_F$.

When a detection occurs in a given gate, the following ($F \Delta t - 1$) gates are removed. After a dead-time is elapsed the gates are reactivated. However, there is still a non-negligible afterpulsing probability, which increases the detection probability in the subsequent gates in the frame, after a dead-time was applied.

\begin{figure}[h!]
\centering
\includegraphics[scale=0.4,trim=0cm 0cm 0cm 0cm,clip]{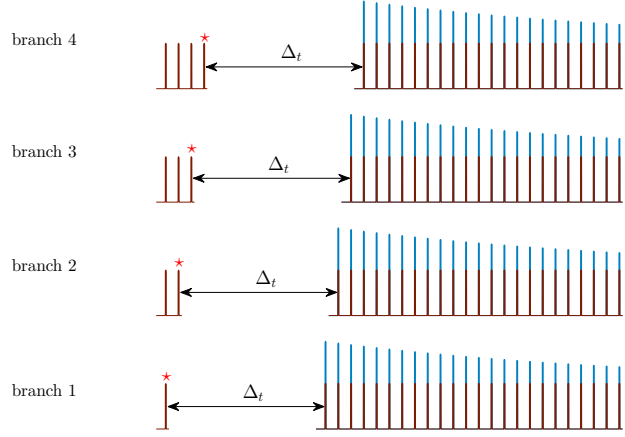}
\caption{Four examples of branch configurations: Detection configuration, where the additional afterpulsing contribution (blue component) on branch-$l$ arises from a detection event, and increments the detection probabilities (brown component). The detection event is marked with a red-star on the respective gate. The gap among the pulses corresponds to a dead-time.}\label{F:A}
\end{figure}

To perform the calculations, it is necessary to identify all possible detection configurations. In Figure~\ref{F:A}, we depict a possible detection configuration, and refer to it as a branch-$l$. The index $l$ is the number of gate where the detection event occurs. Then, we calculate the average afterpulsing contribution on each branch-$l$ as follows,
\begin{eqnarray}
p_{afa,l}&=&\frac{kQ\mathrm{e}^{-k\Delta_t}}{N-l}\sum_{m=0}^{N-l-1}\rho^{-m}\nonumber\\
&=&\frac{kQ\mathrm{e}^{-k\Delta_t}}{(N-l)(1-\rho)}\left[1-\rho^{-(N-l)}\right],
\end{eqnarray}
where $\rho=\mathrm{e}^{-kt_F}$, and $N=(t_S-\Delta_t)F$.

\noindent Since for branch-$l$, $N-l$ gates get contribution from the afterpulsing, we average over all branches $p_{afa,l}$, as follows, 
\begin{eqnarray}
p_{af}'&=&\frac{\sum_{l=0}^{N-1}(N-l)p_{afa,l}}{N_{Ta}}\nonumber \\
&=&\frac{kQ\mathrm{e}^{-k\Delta_t}}{N_a(1-\rho)}\left[1-\frac{\rho(1-\rho^{-N})}{N(1-\rho)}\right],
\end{eqnarray}
where $N_{Ta}=N_aN$ and $N_a=(N+1)/2$. 

\section{SPDs calibration} \label{Ap:B}
We use our model to fit the values the internal detector parameters of an id201 from idQuantique, an APD-based single-photon detector (SPD), using these values  to perform all calculations in this section. The values of its internal detectors parameters ($\eta$, $p_{dc}$, $Q$ and $\tau$) depend on the voltage and temperature settings. Once the voltage setting is fixed to choose the efficiency, the other internal parameters are also modified. The afterpulsing's parameters ($Q$ and $\tau$) exhibit effective values describing the internal carrier-traps assembly. Additionally, they depend on the dead-time setting value, which reduces the effect of the fast-decay carrier-traps.  \\
\begin{figure}[h!]
\centering
\begin{tabular}{cc}
\includegraphics[scale=0.35]{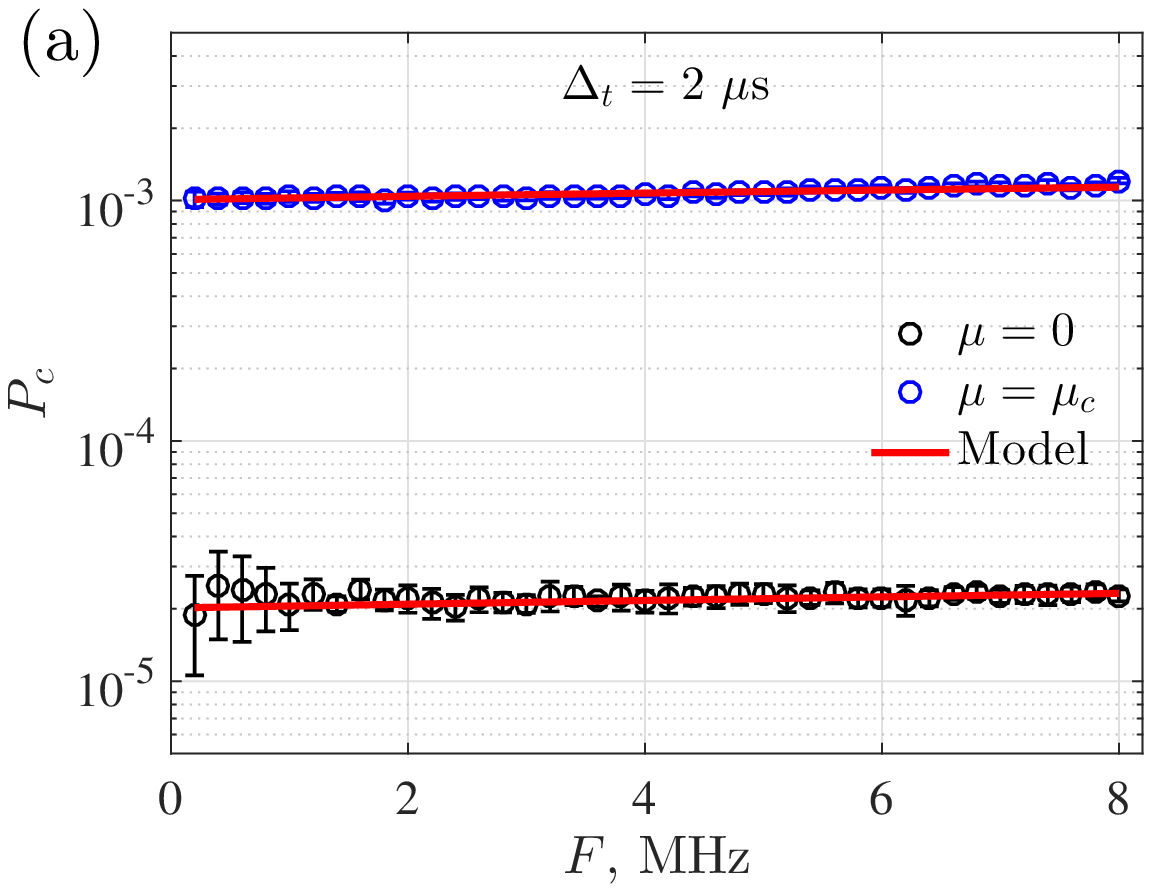}&
\includegraphics[scale=0.35]{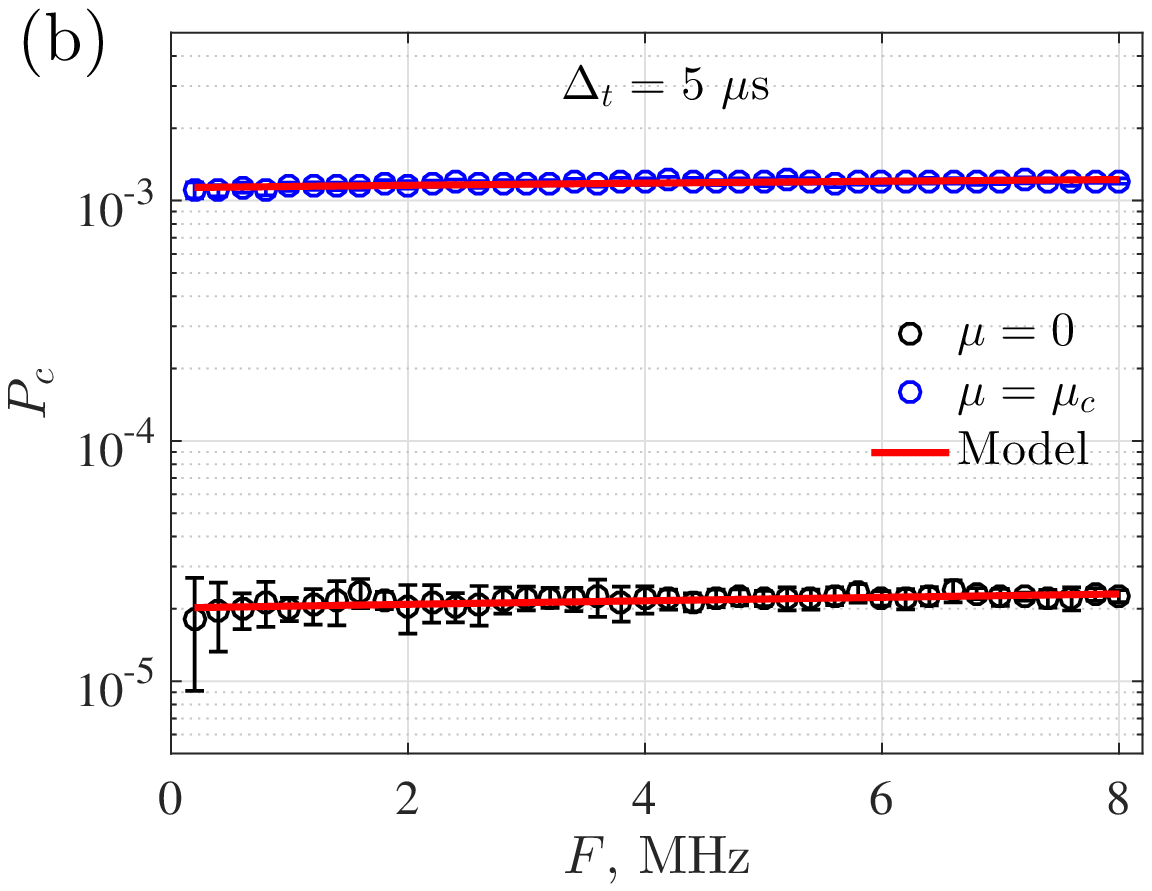}\\
\includegraphics[scale=0.35]{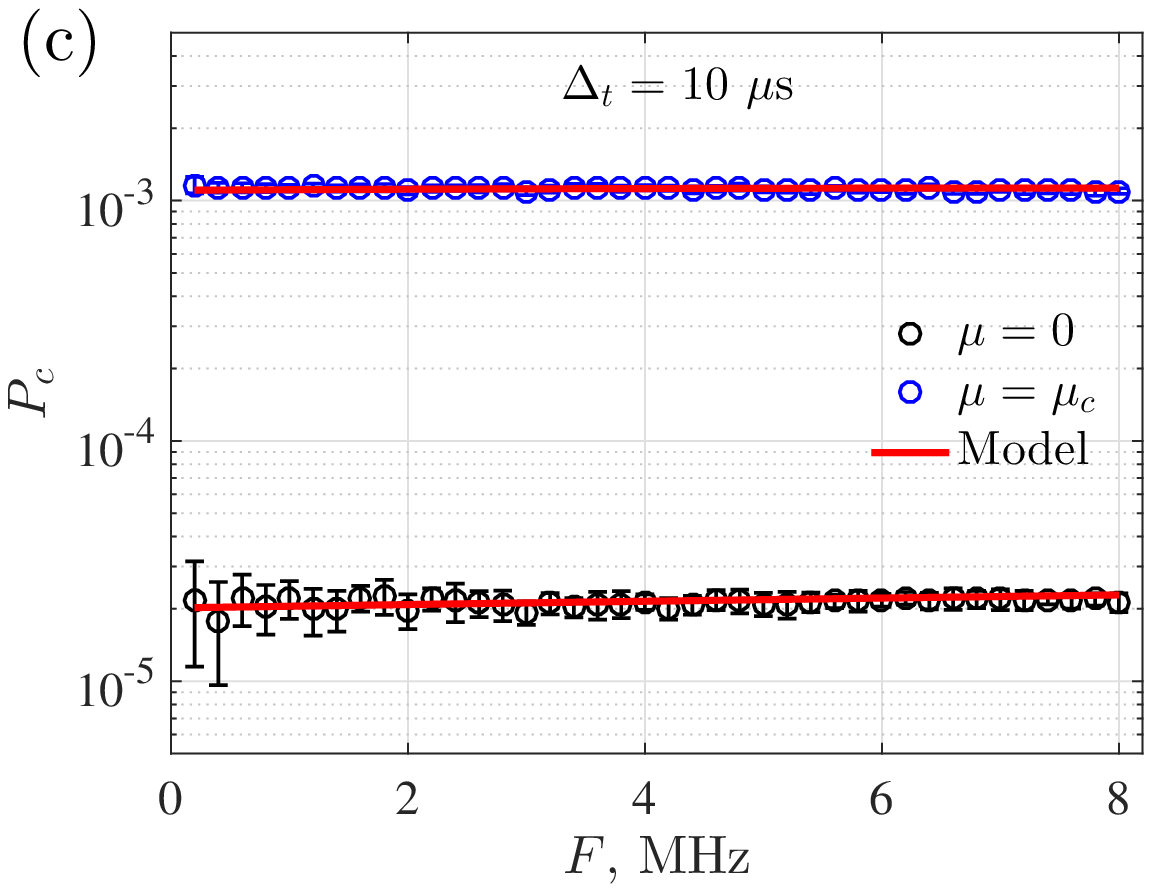}&
\includegraphics[scale=0.35]{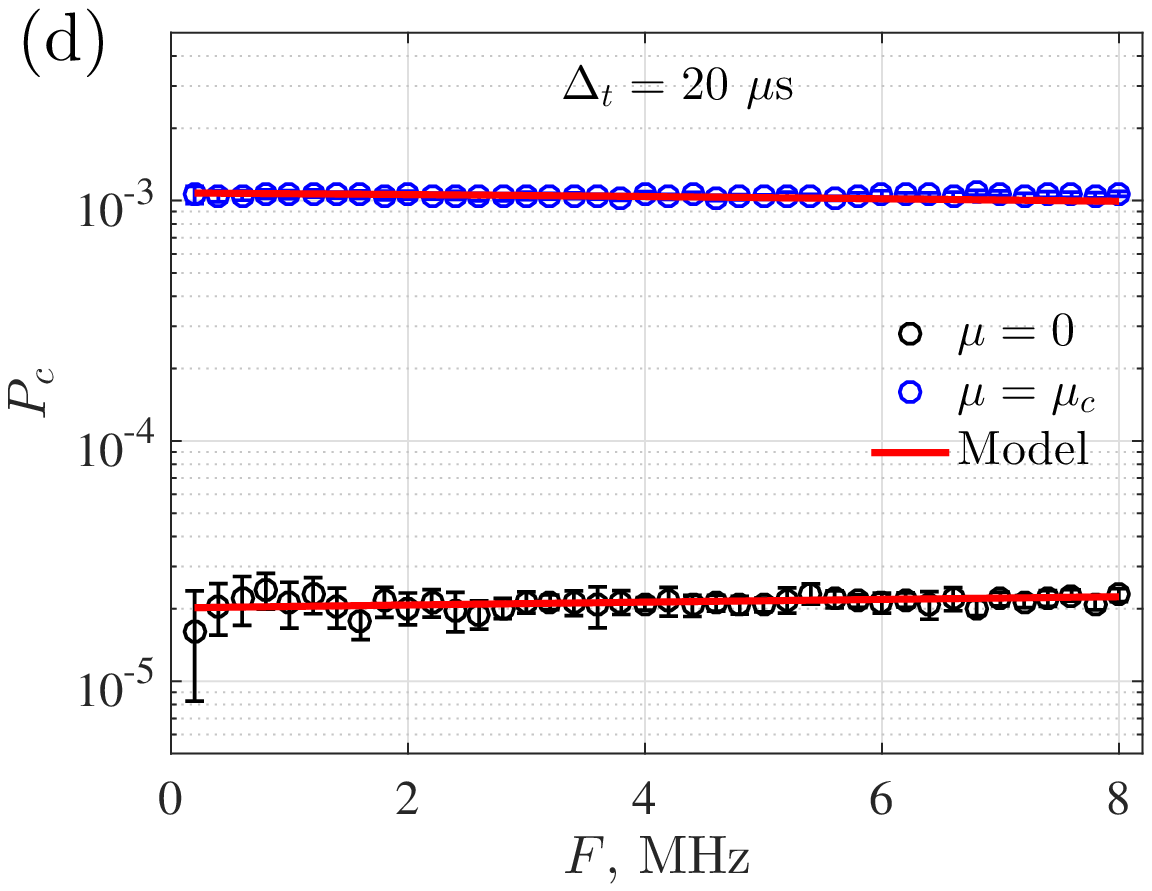}
\end{tabular}
\caption{\small Experimental click probability $P_c$ (blue and black circles) vs model $P_{TC}$ (red solid line) using the fitting values of internal detector parameters  for a id201. Considering gates with a pulse-width of 2.5~ns, and dead-time values: (a) $\Delta_t=2\ \mu$s; (b) $\Delta_t=5\ \mu$s; (c) $\Delta_t=10\ \mu$s; (d) $\Delta_t=20\ \mu$s.}\label{F:B}
\end{figure}

\noindent In order to characterize the detector using a configuration with lowest dark-counts and  afterpulsing contributions, we use the allowed frequency range, limited by the internal electronics of the id201. Also we chose four dead-times values, $\Delta_t\in\{2\ \mu\mathrm{s},\ 5\ \mu\mathrm{s},\ 10\ \mu\mathrm{s} ,\ 20\ \mu\mathrm{s}\}$, and $T_S=T_{fr}=1.00$~s. In Fig.~\ref{F:B}, we show the results and their respective fittings using the model of $P_{TC}$ described in Section~\ref{Sec:2}, removing the SPDs' identifier ($w$) since we use only one detector. Here, we use a single source, $P_{ph0}=\exp(-\eta\mu)$. Denoting $\mu$ as the mean photon-number at the APD, with values $\{0,\ \mu_c\}$. Corresponding a states without light and with a coherent state with $\mu_c=(1.16\pm0.06)\times 10^{-2}$. The values of the internal detector parameters are reported in Table~\ref{T:1}. 
\end{appendix}

\bigskip
\noindent Consejo Nacional de Ciencia y Tecnología (CONACyT): (\textbf{PDCPN 2015-624}), Proyecto Insignia de la Universidad de Guanajuato.

\bibliography{afterPQKDb}

\end{document}